\documentclass[english,10pt]{article}

\usepackage[latin9]{inputenc}
\usepackage[margin=1.35in]{geometry}
\usepackage[parfill]{parskip}

    \usepackage{amsfonts, amsmath, amsthm, array, booktabs, bm, centernot, dcolumn, dsfont, endnotes, enumerate, epigraph, float, geometry, graphicx, makecell, mathtools, multicol, nicefrac, pbox, pdflscape, rotating, soul, subcaption, subfiles, titling, todonotes, txfonts, type1cm, verbatim}
    \usepackage[titletoc,title]{appendix}
    \usepackage[en-US]{datetime2}

    \usepackage[longnamesfirst,semicolon,elide]{natbib}
    \setcitestyle{authoryear,open={(},close={)}}

    \usepackage[labelfont=bf]{caption}
	\usepackage{setspace}
	\usepackage{chngcntr}

	\usepackage[shortlabels]{enumitem}
	\setlist[1]{itemsep=-0.25em}

	\definecolor{darker-blue}{HTML}{0D47A1}
    \definecolor{table-gray}{gray}{0.8}

    \usepackage[multiple, hang, flushmargin]{footmisc}
    \usepackage{fnpct}
    \usepackage{hyperref}
    \hypersetup{colorlinks=true, linkcolor=darker-blue, citecolor=darker-blue, urlcolor=darker-blue}
    \usepackage{footnotebackref}

	\usepackage[T1]{fontenc}
	\usepackage{charter}
	\usepackage[activate={true,nocompatibility},final,tracking=true,kerning=true,spacing=true,factor=1100,stretch=10,shrink=10]{microtype}
	\microtypecontext{spacing=nonfrench}

	\newcommand{\ra}[1]{\renewcommand{\arraystretch}{#1}}
	\newcolumntype{d}[1]{D{.}{.}{#1}}

    \newcommand{\ie}{\textit{i.e.}, }
    \newcommand{\eg}{\textit{e.g.}, }
    
    \newcommand{\expect}{\operatorname{E}\expectarg}
    \DeclarePairedDelimiterX{\expectarg}[1]{[}{]}{%
      \ifnum\currentgrouptype=16 \else\begingroup\fi
      \activatebar#1
      \ifnum\currentgrouptype=16 \else\endgroup\fi
    }
    \newcommand{\innermid}{\nonscript\;\delimsize\vert\nonscript\;}
    \newcommand{\activatebar}{%
      \begingroup\lccode`\~=`\|
      \lowercase{\endgroup\let~}\innermid 
      \mathcode`|=\string"8000
    }
    
\begin{document}


\title{
    Machine learning the first stage in 2SLS: Practical guidance from bias decomposition and simulation\bigskip
}
\author{~
  Connor Lennon \hspace*{1em}
  Edward Rubin\thanks{\footnotesize Lennon: Tacoma Power (\href{mailto:clennon@uoregon.edu}{clennon@uoregon.edu}). Rubin (corresponding author): University of Oregon, Department of Economics, 1285 University of Oregon, 175 Prince Lucien Campbell Hall, Eugene, Oregon 97403-1205 (\href{mailto:edwardr@uoregon.edu}{edwardr@uoregon.edu}). Waddell: University of Oregon, Department of Economics and Research Fellow, IZA Bonn (\href{mailto:waddell@uoregon.edu}{waddell@uoregon.edu}). We thank Joshua Angrist, Daniel Chen, Jiafeng Chen, Jonathan Davis, Thomas Dee, Max Farrel, Brigham Frandsen, Benjamin Hansen, Greg Lewis, John Loeser, Grant McDermott, Douglas Miller, James Stock, Jeffrey Wooldridge, and Eric Zou for helpful discussion. Lennon was affiliated with the University of Oregon during most of the project's work. The project's scripts and data will be posted to a publicly available Github repository. The project is exempt from IRB, as it does not involve human subjects. The authors have nothing to disclose.} \hspace*{1em}
  {Glen R. Waddell}
}


\begin{titlingpage}
  \setlength{\droptitle}{0pt}
    \date{\today}
  \maketitle
  \begin{abstract}
  \singlespacing
  \normalsize
  \noindent Machine learning (ML) primarily evolved to solve ``prediction problems.'' The first stage of two-stage least squares (2SLS) is a prediction problem, suggesting potential gains from ML first-stage assistance. However, little guidance exists on when ML helps 2SLS---or when it hurts. We investigate the implications of inserting ML into 2SLS, decomposing the bias into three informative components. Mechanically, ML-in-2SLS procedures face issues common to prediction \textit{and} causal-inference settings---and their interaction. Through simulation, we show linear ML methods (\eg post-Lasso) work well, while nonlinear methods (\eg random forests, neural nets) generate substantial bias in second-stage estimates---potentially \textit{exceeding} the bias of endogenous OLS.
  
  \vspace{1em}
  
  
 
  \end{abstract}
\end{titlingpage}

\onehalfspacing

\section{Introduction}




Machine learning (ML) methods appear in diverse empirical econometric applications. Despite this enthusiasm, the literature has little to say about the practical implications of combining ML methods with two-stage least squares (2SLS). In this paper we provide practical guidance on the potential risks and benefits of inserting standard ML methods into the first stage of 2SLS---drawing insights from a decomposition of potential sources of bias and a series of Monte Carlo simulations. The decomposition shows that ML-in-2SLS overlaps with the canonical \textit{forbidden regression} but also reveals additional potential sources of bias. Our simulation results illustrate this potential is legitimate: some non-linear ML approaches generate larger second-stage bias than the original endogenous regression.

The motivation behind integrating machine learning in two-stage least squares is clear: to the extent incorporating ``better" first-stage predictions is possible, researchers can obtain more precise second-stage estimates.\footnote{The integration of 2SLS and ML has already appeared in applied work across many fields. Early examples have used linear models in the fist stage---predictors that typically perform well in our decomposition and simulation. For example, estimating labor market impacts of imprisonment \citep{MuellerSmith2015}, the effects of racial-composition shocks during the Great Migration \citep{Derenoncourt2019}, the effect of expropriation on growth \citep{Chen2020}, the ``true'' size of China's GDP growth \citep{Chen2019}, the inter-generational transmission of health \citep{Bevis2020}, and the heterogeneous impacts of family size and parental labor supply \citep{Biewen2020}.} Because most ML methods are built explicitly for prediction---they typically outperform ordinary-least squares (OLS) at this task---using ML for first-stage predictions seems quite natural. \cite{Mullainathan2017} likewise highlight the reasoning that might lead a practitioner to inject ML into the 2SLS framework: ``Machine learning... revolves around prediction" and ``belongs in the part of the toolbox marked $\hat{y}$ rather than in the more familiar $\hat{\beta}$ compartment." The authors immediately recognize that ``the first stage of a linear instrumental variables regression is effectively prediction."\footnote{Similar intuition is offered in \citep{Belloni2011, Belloni2012, Belloni2013, Chernozhukov2015, Chernozhukov2018, Singh2019, Angrist2020, Singh2020, Chen2020-IV} inclusive of new artificial intelligence (AI) methods \citep{Hartford2017,Bennett2020, Liu2020}. In a recent working paper, \cite{Chen2020-IV} also recognizes this motivation, suggesting that the traditional OLS-based implementation of 2SLS ``leaves on the table some variation provided by the instruments that may improve precision of estimates.'' If one is willing to accept the fairly strong assumption that any function (nonlinear or linear) of valid instruments is a valid instrument, \cite{Chen2020-IV} provides an interesting solution to some of the challenges involved with including ML methods in 2SLS. We do not make this assumption.} The risks of adopting out-of-the-box ML methods for 2SLS-type applications are less clear. 

In this paper, we document the risks ML-in-2SLS poses for practitioners of applied econometrics. How does curating and generating first-stage predictions with ML affect the downstream, second-stage causal estimates of two-stage least squares?
\footnote{With valid instruments, applying OLS in the first stage of 2SLS produces predictions $(\hat{x})$ that are a linear combination of the exogenous instruments. Thus, $\hat{x}$ is itself exogenous in the traditional 2SLS procedure. Predictions produced by nonlinear functions are not guaranteed to be orthogonal to their residuals, generating additional bias/inconsistency in second-stage estimates.} Using a simple decomposition, we discuss several phenomena that can bias ML-based 2SLS away from its target parameters. Some of these phenomena are implications of the \textit{forbidden regression} \citep{Angrist2001,Angrist2009,Wooldridge2010}, which na{\"i}ve implementations of ML in 2SLS are likely to lead to---injecting the predictions of a nonlinear estimator into the first stage of 2SLS.\footnote{Another flavor of the \textit{forbidden regression} involves applying different specifications of controls in the first and second stages. Most out-of-the-box ML methods do not offer a method to ensure that second-stage controls are used for prediction in the ML-based first stage (and \textit{in the correct functional form}). There are \textit{ad hoc} solutions to this problem---writing custom functions that implement the ML algorithm \textit{plus} a linear specification of the controls/fixed effects, or residualizing (i.e., Frisch-Waugh-Lovell). For an example, see the \href{https://github.com/lrberge/fixest}{\texttt{fixest}} package in \texttt{R} and its \texttt{feNmlm()} function, which is written to efficiently estimate maximum likelihood models with multiple fixed-effect (\ie large factor variables). This issue is particularly important for situations where conditioning on controls/fixed effects is integral to the instruments' exogeneity. Again, ML methods will, in this way, expose researchers to potential pitfalls.} 


Other issues are less common to ``traditional" econometrics but become key, we argue, to understanding ML-based results. These include: 
\begin{itemize}
	\item \textbf{Recovering endogeneity:} If the prediction algorithm is \textit{too} good, then the first-stage predictor may entirely recover the endogenous regressor (including both \textit{good} and \textit{bad} variation). With (\textit{i}) a small set of valid instruments and (\textit{ii}) a linear estimator (\eg OLS), this scenario is of less concern. As the number of potential instruments increases and the estimator becomes more nonlinear and flexible (a hallmark of many ML methods), we show that this concern becomes real.
	\item \textbf{Exclusion restrictions:} ML methods are not designed to choose exclusion restrictions. If one relies on ML methods to determine a nonlinear functional form, choose instruments, and select first-stage controls in a 2SLS framework, then one must ultimately assume that the algorithm is capable of settling on a valid exclusion restriction. This places a lot of trust in ML to do something it is not typically designed to do. As \cite{Angrist2020} point out, nonlinear estimators generate nonlinear combinations of the original instruments and thereby require additional exclusion restrictions \textit{beyond} the original exclusion restriction implied by the linear combination of the instruments. With highly flexible ML methods, the set of exclusion restrictions is nearly infinite. One must either assume that (\textit{i}) the ML algorithm will choose the appropriate exclusion restrictions or (\textit{ii}) {\it all} possible exclusion restrictions are valid (as the algorithm's choice set is infinite). 
	\item \textbf{Amplified bias:} As we show below, the bias of second-stage estimates in 2SLS is inversely related to the variance of the first-stage predictions $(\hat{x})$. Most ML methods reduce variance in the predictions (to reduce out-of-sample prediction performance in the canonical \textit{bias-variance tradeoff}). This variance-reduction strategy leads to inflated bias in second-stage applications---a consideration not typical to OLS-based 2SLS applications.
\end{itemize}

In our simulations, most ML-rooted solutions that use common ML procedures in the first stage of 2SLS fail to improve upon standard 2SLS (\ie using OLS in the first stage) and generate more bias. Two linear estimators are the exception: post-Lasso selection and principal component analysis (PCA). Post-Lasso and PCA perform at least as well as standard OLS-based 2SLS. Perhaps more importantly, we show that highly nonlinear tree-based methods (\eg random forests and boosted trees) can amplify bias, providing parameter estimates farther from \textit{truth} than na\"ive OLS regressions that ignore endogeneity. Given sufficient training time, na\"ive implementations of neural networks in 2SLS can reproduce the original OLS bias, with little to no advantage over traditional approaches to recovering exogenous identifying variation through 2SLS.\footnote{This ignores the practical as well. With our resources, the simulation for neural network frequently took several hours to complete which is considerably longer than the time it takes to run a traditional 2SLS.}

In Section~\ref{sec:models} we formalize the theoretical settings and define the estimators. In Section~\ref{sec:dgp} we introduce two data-generating processes. Because practitioners' use cases differ, we compare two general cases: one simple (and dense) case and a second, more complex and sparse case. In Section~\ref{sec:results} we present the empirical results for the discussed estimators and DGPs. In Section~\ref{sec:results} we also build a theoretical decomposition that explain how and why ML-based 2SLS procedures might increase second-stage bias. 

While this paper focuses on a practical, \textit{ad hoc} approach of inserting ML into the first stage of 2SLS, where ML methods might be thought of as complimenting traditional approaches to causal estimation, a separate strand of the literature offers tailor-made ML-based estimators in 2SLS-like structures. If one is willing to depart from a traditional 2SLS structure and accept different (typically stronger) identifying assumptions, these methods potentially capture more of the available ``first-stage'' variation without suffering from the issues we highlight in this paper. In Section~\ref{sec:discussion} we discuss such a solution to the problems inherent to the \textit{ad hoc} ML-based 2SLS approach. In Section~\ref{sec:conclussion} we offer concluding remarks.


Ultimately we conclude that while ML methods offer many promises for a range of applications, most out-of-the-box ML methods are not well suited for the first stage of two-stage least squares. Moreover, applying the \textit{wrong} ML method in the first stage can actually generate more bias in parameter estimates than entirely ignoring endogeneity.

\section{Models} \label{sec:models}

\subsection{The problem}

Applied researchers commonly apply 2SLS to estimate the causal effect of some $x$ on some $y$ in a setting where the exogeneity of $x$ cannot reasonably be assumed. In other words, where 
\begin{equation}
    y = \beta_0 + \beta_1 x + u~,
    \label{eq:pop0}
\end{equation}
there is concern over the potential for non-zero covariance between the variable of interest $x$ and the disturbance $u$ when estimating the parameter $\beta_1$.

Let $\mathbf{z}$ denote a vector of \textit{instrumental variables}. We express the first stage of a 2SLS estimates $x$ as a function of these instruments:
\begin{equation}
  x = f(\mathbf{z}) + \varepsilon~.
  \label{eq:pop1}
\end{equation}
In its traditional OLS-based implementation, $f(\mathbf{z})$ is linear in $\mathbf{z}$.

Defining the predictions from (\ref{eq:pop1}) as $\hat{x} = f(\mathbf{z})$, the second stage of the 2SLS procedure then regresses the outcome variable $y$ on $\hat{x}$,
\begin{equation}
  y = \gamma_0 + \gamma_1 \hat{x} + w~,
\end{equation}
to achieve an estimate for $\beta_1$ in (\ref{eq:pop0}). We define $\hat{\gamma}_1$ as this estimate of $\beta_1$. If the instruments are valid (\ie predictive of $x$ and uncorrelated with $u$) and $\hat{x}$ results from an OLS regression, then $\hat{x}$ will also be exogenous.\footnote{We assume homogeneous treatment effects, which removes the requirement of monotonicity.} The second stage of OLS-implemented 2SLS then generates consistent estimates of $\beta_1$, interpreted as the causal effect of $x$ on $y$.
 
So why introduce ML? Applications of 2SLS identify the effect of $x$ on $y$ by extracting only a fraction of the ``good'' (exogenous) variation in $x$. The hope for ML-based 2SLS methods is that researchers can extract more of the good variation in $x$---a more flexible fit of the exogenous variation---while still omitting the bad variation. This desire has likely increased following \cite{lee2020}, who argue that many traditional evaluations of instrumental variables considerably overestimate their significance.

\subsection{Estimators}

In the analysis below we examine three classes of 2SLS-motivated estimators:

\textbf{Class 1: ``Traditional" two-stage regression methods:} This set of estimators covers the standard two-stage regression estimators in an econometrician's toolbox: two-stage least squares, (unbiased) split-sample IV \citep{Angrist1995}, the Fuller implementation of limited-information maximum likelihood (LIML) \citep{Anderson1949,Fuller1977}, and jackknife IV (JIVE) \citep{Angrist1999}. These methods overlap in three important ways: they (\textit{i}) employ a two-stage approach  (\textit{ii}) whose first stage creates a linear combination of the instruments (\textit{iii}) with no formal variable selection.

\textbf{Class 2: Machine-curated variable selections in standard 2SLS:} This second class augments the standard OLS-based version of 2SLS with variable selection/synthesis. Specifically, these methods feature an additional procedure, \textit{prior to the first stage}, that downselects or combines $\mathbf{z}$ into a more parsimonious set of variables. The elements of this more parsimonious expression of $\mathbf{z}$ then appear in the first stage. The rest of the 2SLS process proceeds as usual (\ie OLS). Importantly, while these models feature variable selection or synthesis, they also preserve linearity in both stages. Because these estimates result from linear combinations of $\mathbf{z}$, the original exclusion restriction of $\mathbf{z}$ passes through to the selected/synthesized instruments.

Our first machine-curated method is the post-Lasso procedure of \cite{Belloni2012}, which first estimates the linear relationship between $x$ and $\mathbf{z}$ (a linearized version of \ref{eq:pop1}) using penalized regression. This penalized regression minimizes the sum of squared error (SSE) \textit{plus} a penalty proportional to the sum of the coefficients' magnitudes. That is, $\lambda \times \| \bm{\gamma} \|$, where $\bm{\gamma}$ is the vector of coefficients on the (standardized) instruments and $\lambda$ is a the shrinkage parameter chosen by the researcher (typically via cross validation). Because each instrument's coefficient-based penalty changes discontinuously when moving away from $\gamma_i=0$, Lasso can be used to select a set of \textit{stronger} instruments (whose coefficients are non-zero). Post-Lasso selects the instruments whose coefficients are non-zero and then estimates standard, OLS-based 2SLS using those selected instruments.\footnote{\cite{Angrist2020} notes that this methodology may suffer from potentially unseen pre-test bias. Because our model comes from relatively strong instruments, as with the intuition of \cite{Zhao2020}, we do not estimate de-biased Lasso models. We therefore allow post-Lasso to serve as a representation of both.}

Principal-component analysis (PCA) offers an alternative route to simplifying $\mathbf{z}$ by selecting $\mathbf{z}$'s first $k$ principal components \citep{Pearson1901}. Thus, as the second machine-curated method we consider, principal-component analysis (PCA) applied to 2SLS (as in \cite{Ng2009} and \cite{Winkelried2011}). This approach passes a set of principal components into the first stage of standard OLS-based 2SLS. While PCA may reduce the first stage's interpretability, this approach can drastically reduce the number of first-stage instruments while retaining considerable explanatory power.

\textbf{Class 3: ML-based first stages in 2SLS:} Our final class of estimators retains the general two-step framework of 2SLS but replaces the first stage with a variety of cross-validated ML algorithms. We evaluate a meaningful subset of machine-learning methods suitable for regression, including random forest \citep{Ho1995,Breiman2001}, boosted trees \citep{Breiman1997,Mason1999,Friedman2001,Friedman2002}, neural networks \citep{Turing1948,Mcculloch1943,Farley1954}, and Lasso \citep{Tibshirani1996,Santosa1986}.\footnote{For our purposes, the contributions of \cite{Srivastava2014} (dropout), \cite{Ioffe2015} (batch normalization), and \cite{Kingma2017} (stochastic optimization) are particularly relevant.}\footnote{For a nice review of ML methods in applied economics, including Lasso, tree-based methods, and neural networks, please see \cite{Storm2019}. For  broader and more in-depth coverage (from the authors of many of the methods), see \cite{James2013} and \cite{Hastie2009}.} Notably, most of these algorithms offer considerable flexibility (\eg nonlinearity in $\mathbf{z}$) and variable selection (to varying degrees). This class offers considerable insights into the merits of off-the-shelf ML methods for machine-assisted 2SLS.

\section{Data-generating processes} \label{sec:dgp}

In order to examine the performance of ML in the predictive stage of 2SLS---in absolute terms and relative to ``traditional'' options---we employ two general data-generating processes (DGPs). For reasons described below we refer to the two DGPs as the \textit{low-complexity} case and the \textit{high-complexity} case. In practice, the researcher rarely knows the extent to which her case is \textit{complex}, particularly in terms of extent of nonlinearity or the efficient number of instruments. While ``complexity" is certainly subjective, our intention is to bookend the settings for which an applied researcher might apply ML-based 2SLS.
 
\subsection{A \textit{low-complexity} case}

In this case, we aim to depict the performance of various estimators when the DGP is simple (few instruments and non-sparse) and closely matches the ideal scenario for OLS-based 2SLS: an endogenous regressor that is a linear combination of a relatively small set of strongly predictive, exogenous instruments. This case is applicable to researchers seeking to estimate the causal effect of a variable of interest $x_1$ on outcome $y$,
\begin{equation}
    y = \beta_0 + \beta_1 x_1 + \varepsilon_y~,
    \label{eq:low-pop}
\end{equation}
but facing the challenge (e.g., omitted variables, simultaneity) that $x_1$ is endogenous and $\expect*{\varepsilon_y \big| x_1}\neq0$ prevents OLS from cleanly identifying $\beta_1$ in (\ref{eq:low-pop}). Importantly, the causal effect $\beta_1$ is common across all individuals, which ensures differences across estimators are not due to the estimators recovering different local average treatment effects (LATEs).

In this low-complexity scenario, ML-based 2SLS methods are overkill: neither variable selection nor nonlinearity are necessary. In fact, our results demonstrate that ML methods can increase bias relative to 2SLS, even relative to endogenous OLS.

Formally, to model a scenario with a single endogenous regressor $(x_1)$ and a small set of valid (and {\it individually} strong) instruments, we define the DGP as
\begin{align*}
    \varepsilon_y &=  \beta_2 x_2 + \eta~,
    \\
    x_2 &= 1 + \varepsilon_c~,~ \text{and}
    \\
    x_1 &= g_x(\mathbf{z}) + \varepsilon_c~,
\end{align*}
drawing special attention to the inclusion of $\varepsilon_c$ as the disturbance common to both $x_1$ (the variable of interest) and $x_2$ (the \textit{omitted} variable). This common error follows a standard normal distribution; $\eta$ is distributed uniformly between $-1$ and $1$.

We assume that a set of valid instruments $\mathbf{z}$ exists such that $\expect*{\varepsilon_y | \mathbf{z}} = 0$ and $\expect*{x_1 | \mathbf{z}} \neq 0$ (we focus on the case where $|\mathbf{z}|=7$). We also anticipate that the researcher has no beliefs or insights about the functional form of $g_x(\cdot)$, as is often the case in practice. In the true DGP for this case, $g_{x}(\mathbf{z}) = \sum_{i=1}^{7} z_i$. That is, $g_{x}(\cdot)$ is linear. 

In particular, we draw the instruments $\mathbf{z}$ from a multivariate normal distribution centered at zero (\ie $\expect*{\mathbf{z}} = \mathbf{0}$) with variance-covariance matrix $\Sigma_\mathbf{z}$ where $\widehat{\text{Cov}}(z_i,\,z_j) = 0.6^{|h-k|}$ (and thus $\text{Var}(z_i) = 1$ for each $i$). By implication, $x_1 \sim N(0,\, \text{Grand Sum}(\Sigma_\mathbf{z}) + 1)$.

In full, then, the data represents the following system of equations:
\begin{align*}
    y~ &= \underset{(=1)}{\beta_0} + \underset{(=1)}{\beta_1} x_1 + \underset{(=1)}{\beta_2} x_2 + \eta~,
    \\
    x_2 &= 1 + \varepsilon_s~,
    \\
    x_1 &= g_x(\mathbf{z}, \varepsilon_s) = \sum^7_{i=1} z_i + \varepsilon_s~.
\end{align*}

Notably, the specification of the instruments in this DGP produces a very strong first-stage with a relatively large \textit{concentration parameter} \citep{Belloni2012}. Put simply, the concentration parameter $\mu^2$ describes the extent to which the weak-instrument problem may arise within a given DGP. A higher value of $\mu^2$ implies that 2SLS, without variable selection, will converge to the true $\beta_1$ at relatively small sample sizes.\footnote{In this ``low-complexit'' case, $\mu^2 \approx n\times 20.71$, which exceeds the values in \cite{Belloni2012}. We discuss $\mu^2$'s role further in the \textit{high-complexity case} section below.} Consequently, the low-complexity case allows us to test how machine-curated first stages perform when there is little to be gained from variable selection/synthesis.
    
\subsection{A \textit{high-complexity} case}

As our high-complexity case, we follow \cite{Belloni2012}'s DPG with two extensions. This DGP allows the researcher to tailor instruments' strengths with \textit{many} instruments. Following \cite{Belloni2012}, the DGP in our high-complexity case results from
\begin{align*}
  y &= \beta_0 + \beta_1 x_{1} + \varepsilon_y~,
  \\
  x_{1} &= \bm{\pi} \mathbf{z} + \varepsilon_v~,
 \end{align*}
where
\begin{align*}
  (\varepsilon_y,\, \varepsilon_v) &\sim N\!\left(0,\, \left[
  \begin{matrix}
    \sigma^2_y & \sigma_y \sigma_v\\
    \sigma_v \sigma_y & \sigma^2_v
  \end{matrix}\right] \right)~,
  \\
  \mathbf{z} &= \left[\begin{matrix} z_{1} & z_{2} & \cdots & z_{100} \end{matrix}\right] \sim N(\mathbf{0},\, \Sigma_{z})~,
  \\
  \Sigma_{z}{[j,j]} &= \text{Var}(z_{j}) = \sigma^2_j = 1, \enspace \forall j \in \{1,\, \dots,\, 100\}~, and
  \\
  \Sigma_{z}{[j,k]} &= \widehat{\text{Cov}}(z_{j},\, z_{k}) = \text{Cor}(z_{j},\, z_{k}) = 0.6^{|j-k|}, \enspace \forall (j,\,k) \in \{1,\ \dots,\, 100\}~.
 \end{align*}
As before, we imagine the researcher's interest in this case focuses on identifying $\beta_1$. However, unlike the earlier DGP, the high-complexity case produces sets of relevant and exogenous instruments that vary in their correlation and individual strength (\ie $\pi_i$).

\begin{landscape}
\begin{figure}
    \caption{\textbf{The ``high-complexity'' DGP:} 100 instruments of varying strength}
    \label{fig:dgp-complex}
    \centering
    \caption*{\normalsize \textbf{Order of instruments' coefficients} ($\bm{\pi}$)}
    \begin{subfigure}{0.33\linewidth}
        \centering
        \caption{\textit{Subcase 1:} `Shuffled' coefficients}
        \label{fig:dgp-coef-1}
        \includegraphics[width=0.99\linewidth]{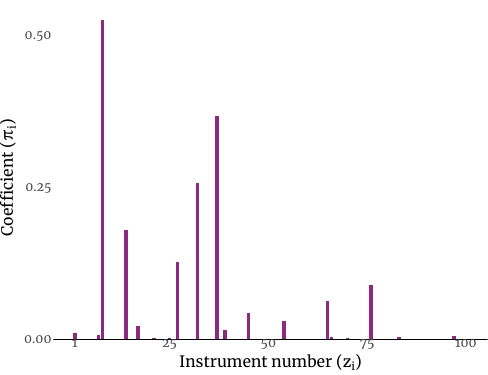}
    \end{subfigure}%
    \begin{subfigure}{0.33\linewidth}
        \centering
        \caption{\textit{Subcase 2:} Coefficients decline from $z_1$}
        \label{fig:dgp-coef-2}
        \includegraphics[width=0.99\linewidth]{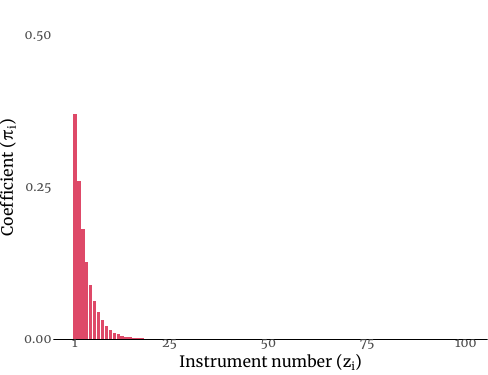}
    \end{subfigure}%
    \begin{subfigure}{0.33\linewidth}
        \centering
        \caption{\textit{Subcase 3:} Coefficients decline from $z_{50}$}
        \label{fig:dgp-coef-3}
        \includegraphics[width=0.99\linewidth]{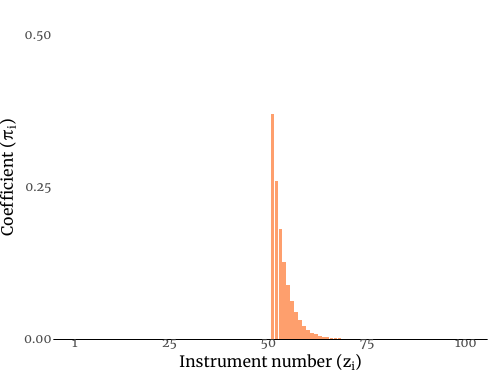}
    \end{subfigure}%
    \hfill \bigskip
    \caption*{\normalsize \textbf{Instruments' correlation} ($\Sigma_z$)}
    \begin{subfigure}{0.33\linewidth}
        \centering
        \caption{Correlation between the 100 instruments}
        \label{fig:dgp-cor-mat-complex}
        \includegraphics[width=0.99\linewidth]{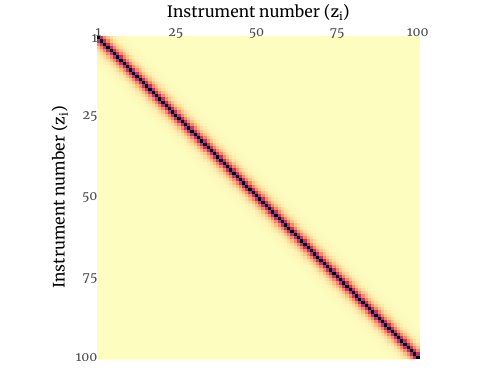}
    \end{subfigure}%
    \begin{subfigure}{0.33\linewidth}
        \centering
        \caption{Correlation of $z_1$ and $z_{50}$ with other instruments}
        \label{fig:dgp-cor-line-complex}
        \includegraphics[width=0.99\linewidth]{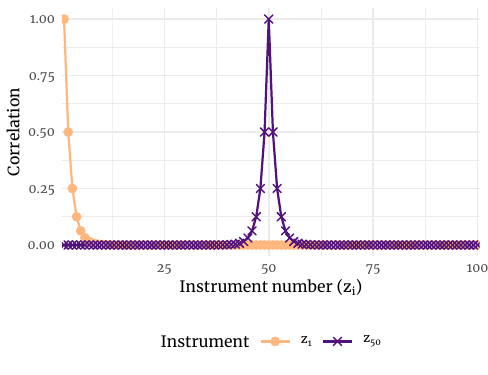}
    \end{subfigure}%
    \hfill
    \caption*{\small The top panels---\ref{fig:dgp-coef-1}, \ref{fig:dgp-coef-2}, and \ref{fig:dgp-coef-3}---illustrate the instruments' coefficients ($\bm{\pi}$) in the DGP for the first stage of each of the three subcases (\ie how the instruments $\mathbf{z}$ relate to $x$ in each subcase). While the three subcases differ in their first-stage coefficients, they share the same variance-covariance structure $\Sigma_z$, which \ref{fig:dgp-cor-mat-complex} depicts ($\text{Var}(z_i)=1,\, \forall i$). Figure~\ref{fig:dgp-cor-line-complex} individually plots the correlation between two instruments ($z_1$ and $z_{50}$) and all of the other instruments. The instruments $z_1$ and $z_{50}$ are specifically of interest because they are the strongest instruments in subcases 2 and 3, respectively.}
\end{figure}
\end{landscape}

In defining the ``exponential'' design of the \textit{first-stage} coefficient vector $\bm{\pi}$, we follow \cite{Belloni2012}: $\bm{\pi}$ captures a ``beta pattern" $\widetilde{\bm{\pi}} = (0.7^{0},\, 0.7^{1},\, 0.7^{2},\, \ \dots,\, 0.7^{99})$ that is then multiplied by a constant $C$, \ie $\bm{\pi} = C \times \widetilde{\bm{\pi}}.$ The constant $C$ implies a value for the concentration parameter, $\mu^2 = \frac{n \bm{\pi}'\Sigma_z \bm{\pi}}{\sigma^2_v}$.\footnote{For a proof of this statement, see \cite{Belloni2012}.} In panels~\ref{fig:dgp-coef-1}--\ref{fig:dgp-coef-3} (Figure~\ref{fig:dgp-complex}) we illustrate the three beta patterns that we adopt in the ``high-complexity" DGP, generating three subcases of this DGP. As described above, the concentration parameter is useful for determining the behavior of IV estimators. Because we are less interested in the case of weak instruments, we use $\mu^2 = 180$, which creates a strong---though fairly sparse---set of instruments as outlined in \cite{Belloni2012}.\footnote{It is important to select this value thoughtfully. Choosing a $\mu^2$ that is too small will simulate a weak-instruments problem. Choosing a $\mu^2$ that is too large will yield a scenario in which all instruments are ``overpoweringly'' valid, which reduces the effectiveness of selection or dimension-reduction techniques. See \citet{Hansen2008} for additional discussion of $\mu^2$.}

\cite{Belloni2012} arrange the coefficients $\bm{\pi}$ in descending order (\ie $\pi_1 > \pi_2 > \cdots > \pi_{100}$). However, the definition of $\Sigma_z$ implies that ``proximate" instruments are more correlated than ``distant'' instruments (\ie $\text{Cor}(z_i,\,z_{i+1}) > \text{Cor}(z_i,\,z_{i+k})$ for $k>1$). Thus, the DGP of \cite{Belloni2012} ensures the strongest instruments correlate with each other. While this feature may be desirable in many contexts, we remain agnostic with regard to whether the strongest instruments are most correlated with each other or with other instruments. However, this agnosticism requires that we consider three sub-cases, each arising from alternative orderings of the coefficients in $\bm{\pi}$ and the covariance of the respective variables:
\begin{itemize}
    \item \textbf{Randomly shuffled:} After generating the coefficients, we randomly re-order them to break the relationship between instruments' strengths and their covariance ($\Sigma_z$).
  \item \textbf{Descending from} $z_1$: In this subcase, as in \cite{Belloni2012}, $\pi_1 > \pi_2 > \cdots > \pi_{100}$.
  \item \textbf{Descending from} $z_{50}$: Here we modify \cite{Belloni2012} by defining $\pi_{50}$ as the largest coefficient: $\pi_{50} > \pi_{51} > \cdots > \pi_{100} > \pi_{1} > \pi_{2} > \cdots > \pi_{49}$. Because ``proximate'' instruments are correlated in $\Sigma_z$, this subcase implies that the strongest instrument ($z_{50}$) is very correlated both with the second-strongest instrument ($z_{51}$) and with the weakest instrument ($z_{49}$).
\end{itemize}

Finally, we define $\sigma^2_v = \bm{\pi}' \Sigma_{z} \bm{\pi}$ (which forces that $\text{Var}(x_1) = 1$) and $\sigma_y = 1$. In panels \ref{fig:dgp-coef-1}-\ref{fig:dgp-coef-3} of Figure~\ref{fig:dgp-complex} we illustrate the cross-instrument correlations implied by $\Sigma_z$: in Panel~\ref{fig:dgp-cor-mat-complex} we show a correlation matrix among the 100 instruments, and in Panel~\ref{fig:dgp-cor-line-complex} we highlight the correlation of $z_{1}$ and $z_{50}$ to each of the other 100 instruments. Instruments are strongly correlated with their neighbors and weakly correlated with non-neighbors, which limits the information accessible from any single instrument.


\section{Results} \label{sec:results}

We now discuss the results of our simulations. In every simulation, we include an ``oracle model'' that extracts the entirety of the exogeneous component of $x_1$ (perfectly removing endogeneity) and a simple OLS model (where we entirely ignore endogeneity). While one might expect the oracle and plain OLS models to bookend the biases in 2SLS-related models, our simulations demonstrate that they \textit{do not}. That is, inserting machine learning into the first stage can lead to outcomes that are even worse than ignoring endogeneity.

In each case, we are interested in the performances of the estimators in terms of their biases and the precision of estimates. Recall that these estimators include three broad classes: (\textit{i}) traditional methods (OLS-based 2SLS, split-sample IV, LIML, and jackknife IV), (\textit{ii}) machine-curated 2SLS (variable-selection or -curation via post-Lasso and PCA), and (\textit{iii}) 2SLS applications with ML-powered predictions in their first stages (\ie replacing first-stage OLS with either Lasso, boosted trees, random forests, or neural networks). 

\subsection{General results}

\begin{landscape}
\begin{figure}
    \caption{Main results---$\hat{\beta}$ distributions across competing two-stage methods}
    \label{fig:results-densities}
    \centering
    \begin{subfigure}{0.5\linewidth}
        \centering
        \caption{\textbf{\textit{Low-complexity} case:} 7 strong instruments}
        \label{fig:density-coef-f1}
        \includegraphics[width=0.99\linewidth]{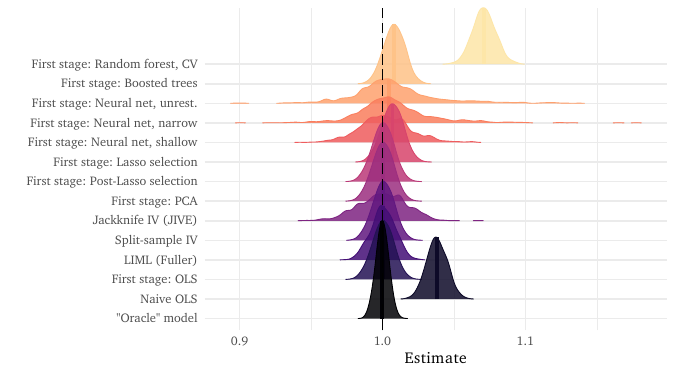}
    \end{subfigure}%
    \begin{subfigure}{0.5\linewidth}
        \centering
        \caption{\textbf{\textit{High-complexity} case 1:} 100 instruments; coefficients `shuffled'}
        \label{fig:density-coef-t1}
        \includegraphics[width=0.99\linewidth]{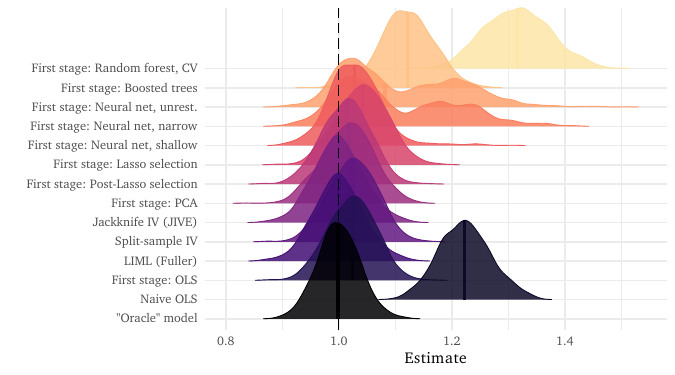}
    \end{subfigure}%
    \hfill
    \begin{subfigure}{0.5\linewidth}
        \centering
        \caption{\textbf{\textit{High-complexity} case 2:} 100 instruments; strength decreases from $z_1$}
        \label{fig:density-coef-t2}
        \includegraphics[width=0.99\linewidth]{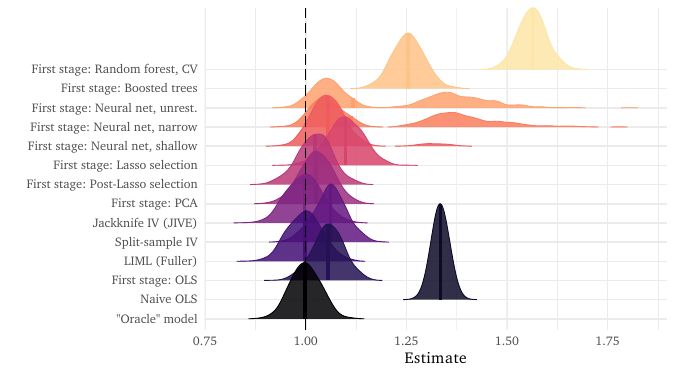}
    \end{subfigure}%
    \begin{subfigure}{0.5\linewidth}
        \centering
        \caption{\textbf{\textit{High-complexity} case 3:} 100 instruments; strength decreases from $z_{50}$}
        \label{fig:density-coef-t3}
        \includegraphics[width=0.99\linewidth]{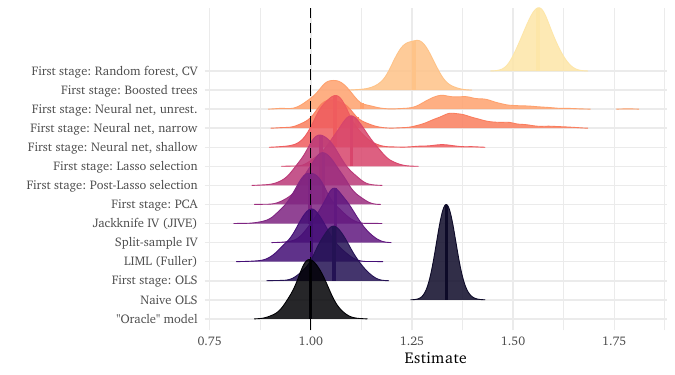}
    \end{subfigure}%
    \caption*{\small Each individual distribution represents the density of estimates from 1,000 iterations of simulation. The true value of the target parameter $\beta$ is 1 (the dashed vertical line). The DGP underlying subfigure~\ref{fig:density-coef-f1} uses 7 strong and exogenous instruments. For subfigures~\ref{fig:density-coef-t1}, \ref{fig:density-coef-t2}, and~\ref{fig:density-coef-t3}, the models have access to 100 exogenous instruments of varying strengths. The three `high-complexity' cases differ in their DGPs' first-stage coefficients but share a common variance structure among the 100 instruments ($\Sigma_z$)---following and extending \cite{Belloni2012}. The \textit{Oracle model} refers to a model using only the exogenous portion of $x$ (the endogenous regressor). In \textit{Naive OLS}, we estimate the parameter without regard for the endogeneity of $x$. The remaining methods each estimate the target parameter with a different variety of approaches toward 2SLS or related methods. Table~\ref{tab:main-results} contains the mean and standard deviation for each distribution.}
\end{figure}
\end{landscape}

In Figure~\ref{fig:results-densities} we depict the distributions of point estimates ($\hat{\beta}_1$) for a given method in the givezn DGPs. Panel~\ref{fig:density-coef-f1} illustrates the low-complexity case; panels~\ref{fig:density-coef-t1}--\ref{fig:density-coef-t3} presents the results for our high-complexity cases. Table~\ref{tab:main-results} summarizes each of these method-by-DGP combinations with the mean and standard error from each. The target parameter $\beta_1$ equals 1 throughout the simulations (indicated with a thin dashed line). Each distribution results from 1,000 iterations of the simulation. 

To those with use-cases that resemble our ``low-complexity case,'' the simulation results have a clear takeaway: PCA-based 2SLS and post-Lasso perform well and offer very safe choices.\footnote{LIML also performs well, but with slightly larger variance. The Jackknife IV estimator yields \textit{very} high variance in this low-complexity DGP, as do Neural Networks.} Important for the practitioner: All four nonlinear ML-in-the-first-stage methods (\ie Lasso, boosted trees, neural networks, and random forests) perform poorly in terms of bias and variance. In fact, random-forest-based 2SLS generates {\it more bias} in $\hat{\beta}_1$ than the OLS estimator that entirely ignores endogeneity---it is possible for an ML-based 2SLS estimator to {\it amplify} bias relative to plain OLS. We discuss the source of this bias amplification in the next section.

In the three high-complexity cases in Table~\ref{tab:main-results} (columns $B$--$D$) and in panels~\ref{fig:density-coef-t1}--\ref{fig:density-coef-t3} of Figure~\ref{fig:results-densities}, LIML and Jackknife IV generate very little bias in their estimates of $\beta_1$, outperforming 2SLS. Across all three DGPs, 2SLS produces mean estimates roughly 2.3--5.8 percent larger than the true parameter, while the centers of LIML's and JIVE's distributions are within 0.4 percent of the true parameter. Injecting random forests into the first stage, on average, produces more biased estimates than na\"ive (endogenous) OLS, generating coefficient estimates that are 32--56 percent larger than the true estimates. Again, one can worsen endogeneity issues by using ML-based 2SLS estimators.



\subsection{Decomposing the bias}
\label{ssec:decomp}
  

To diagnose the sources of bias from different methods, we show one can decompose the wedge between $\beta_1$ and $\hat\beta^{\text{2SLS}}_1$ into three components,
\begin{equation}
\text{Wedge} = \hat{\beta}_{1}^\text{2SLS} - \beta_{1} = f\bigg(
    ~\beta_1 \widehat{\text{Cov}}(\hat{x},e),\,
    \widehat{\text{Cov}}(\hat{x},u),\,
    \frac{1}{\widehat{\text{Var}}(\hat{x})}\bigg)~,
    \label{eq:wedge}
\end{equation}
where $f$ is non-decreasing with respect to each of its arguments, $\hat{x}$ is the first-stage-based prediction of $x$ from some set of valid instruments, $e$ denotes the resulting first-stage residuals $(x - \hat{x})$, and $u$ represents the population disturbance from regressing $y$ on $x$, \ie $u = y - (\beta_0 + \beta_1 x)$. Below we derive and elaborate upon (\ref{eq:wedge}).

Each component of the wedge offers insights into how first-stage methods differentially produce biases---this delivers helpful intuition regarding the pitfalls that may arise in 2SLS applications that include ML-based first stages.

To see the component parts of the bias drawing $\hat\beta^{\text{2SLS}}_1$ away from $\beta_1$, suppose again that the parameter of interest is $\beta_1$, the causal effect of $x$ on $y$ in
\begin{equation}
  y = \beta_0 + \beta_1 x + u~. \label{eq:simple}
\end{equation}
Suppose also that $x$ is endogenous, \ie $\text{Cov}(x,\,u)\neq 0$. The 2SLS estimate of $\beta_1$ comes from estimating
\begin{equation}
  y = \beta_0 + \beta_1 \hat{x} + w~,
  \label{eq:stage2}
\end{equation}
where again $\hat{x}$ is the first-stage-based prediction of $x$ from some set of valid instruments $\mathbf{z} = z_1,\, z_2,\, \ldots,\, z_p$.

Because we estimate the second stage in (\ref{eq:stage2}) via OLS, the estimate for $\beta_1$ can be written
\begin{align}
  \hat{\beta}_{1}{^\text{2SLS}} 
  = \beta_1 + \dfrac{\widehat{\text{Cov}}(\hat{x},w)}{\widehat{\text{Var}}(\hat{x})}~. \label{eq:b-2sls}
\end{align}
where $\widehat{\text{Cov}}(\cdot)$ and $\widehat{\text{Var}}(\cdot)$ refer to the sample-based covariance and variance.

Using (\ref{eq:simple}) and (\ref{eq:stage2}), we can rewrite $w$ as
\begin{align}
  w 
  &= y - \left(\beta_0 + \beta_1 \hat{x}\right) \nonumber \\
  &= \beta_0 + \beta_1 x + u - \beta_0 - \beta_1 \hat{x} \nonumber \\
  &= \beta_1 \left(x - \hat{x} \right) + u \nonumber \\
  &= \beta_1 e + u~, \label{eq:w}
\end{align}
where $e$ is the first-stage residual, the difference between $x$ and $\hat{x}$.

Substituting (\ref{eq:w}) for $w$, we can decompose the covariance in (\ref{eq:b-2sls}) into two components:
\begin{align}
  \widehat{\text{Cov}}(\hat{x},w)
  &= \widehat{\text{Cov}}(\hat{x},\beta_1 e + u) \nonumber \\
  &= \beta_1 \widehat{\text{Cov}}(\hat{x},e) + \widehat{\text{Cov}}(\hat{x},u)~. \label{eq:cov}
\end{align}

If the first-stage predictions $(\hat{x})$ come from OLS, then $\widehat{\text{Cov}}(\hat{x},e)$ is mechanically zero. The second term, $\widehat{\text{Cov}}(\hat{x},u)$, is typically small when $\hat{x}$ comes from a linear combination of valid instruments.

Finally, substituting (\ref{eq:cov}) into (\ref{eq:b-2sls}) yields a helpful expression for the 2SLS estimate for $\beta_1$,
\begin{align}
  \hat{\beta}^\text{2SLS}
  = \beta_1 + \dfrac{\beta_1 \widehat{\text{Cov}}(\hat{x},e) + \widehat{\text{Cov}}(\hat{x},u)}{\widehat{\text{Var}}(\hat{x})}~. \label{eq:b-2sls-wedge}
\end{align}

Again, first-stage OLS guarantees that $\widehat{\text{Cov}}(\hat{x},e)$ is zero and, with valid instruments, that $\widehat{\text{Cov}}(\hat{x},u)$ is small. Whether $\widehat{\text{Var}}(\hat{x})$ is ``small'' is typically of little consequence with OLS (as $\beta_1 \widehat{\text{Cov}}(\hat{x},e) + \widehat{\text{Cov}}(\hat{x},u)$ is typically small). However, all three points can generate important issues when we mix ML methods into the first stage of 2SLS. With ML methods, nothing guarantees that $\widehat{\text{Cov}}(\hat{x},e)$ is zero or that $\widehat{\text{Cov}}(\hat{x},u)$ is small. Moreover, many ML methods are constructed to {\it reduce} the variance of predictions, which further amplifies bias. This variance-reduction aspect is particularly relevant for nonlinear methods.

\subsubsection*{The $\beta_1 \mathrm{Cov}(\hat{x},e)$ component}

For the term $\beta_1 \mathrm{Cov}(\hat{x},e)$ to differ from zero and generate bias, $\beta_1 \neq 0$ and $\mathrm{Cov}(\hat{x},e)\neq 0$. We assume that the population-regression coefficient $\beta_1$ differs from zero. Consequently, the term $\beta_1 \mathrm{Cov}(\hat{x},e)$ only generates bias when $\mathrm{Cov}(\hat{x},e) \neq 0$; $\beta_1$ scales the bias and affects its direction.

By construction, OLS produces predictions that are orthogonal to their residuals, \ie $\widehat{\text{Cov}}(\hat{x},e) = 0.$ This first term is therefore irrelevant when the first stage uses OLS. However, when practitioners adopt other methods in the first stage (\eg non-linear methods) nothing guarantees first-stage predictions are uncorrelated with their residuals. Indeed, this component relates to many researchers' definitions of \textit{the forbidden regression}. This part of the bias results from using estimators whose predictions correlate with their residuals (rather than resulting from a violation of the exclusion restriction). While nonlinear methods can generate $\widehat{\text{Cov}}(\hat{x},e)=0$, many do not (as illustrated by the column $a$ of Table~\ref{tab:bias-decomp}).

In addition, because $\widehat{\text{Cov}}(\hat{x},e)$ typically drops out of OLS regression, OLS-based empirical intuition does not help here. One implication from this non-OLS intuition of $\widehat{\text{Cov}}(\hat{x},e)$ is that the bias generated by this component is proportional to the size of the target parameter $\beta_1$. Where treatment effects are larger, the bias transmitted through this component is also larger.

To understand why some methods produce larger values of $\widehat{\text{Cov}}(\hat{x},e)$ than other methods, first decompose this covariance into $\widehat{\text{Cov}}(\hat{x},x)$ and $\widehat{\text{Var}}(\hat{x})$: 
\begin{align}
    \widehat{\text{Cov}}(\hat{x},e) &= \widehat{\text{Cov}}(\hat{x},\, x-\hat{x}) = \widehat{\text{Cov}}(\hat{x}, x) - \widehat{\text{Var}}(\hat{x})~.
\end{align}

While $\widehat{\text{Cov}}(\hat{x},e)$ is not generally signable, it is bounded between $-\widehat{\text{Var}}(\hat{x})$ and $\widehat{\text{Cov}}(\hat{x},x)$.\footnote{We assume predictions, $\hat{x}$, will have non-negative covariance with the true values, $x$.} Further, we can sign $\beta_1\widehat{\text{Cov}}(\hat{x},e)$ in five subcases:\footnote{We assume $e$, $x$, and $\hat{x}$ have variation and that the predictions $\hat{x}$ positively correlate with the true values $x$.}
\begin{align}
    \text{Sign}\bigg\{~\beta_1 \widehat{\text{Cov}}(\hat{x},e) ~ \bigg\}
    & = \text{Sign} \bigg\{ ~\beta_1 \widehat{\text{Corr}}(\hat{x},e)~ \bigg\}
    \nonumber
    \\ \nonumber
    &= \text{Sign} \bigg\{~\beta_1 \sigma_{e}^{-1}~ \bigg( \widehat{\text{Corr}}(\hat{x},x) \, \sigma_{x} - \sigma_{\hat{x}} \bigg)~ \bigg\}
    \\ \nonumber
    &= \text{Sign} \bigg\{~\beta_1~\bigg\} \, \cdot \, \text{Sign} \bigg\{ ~\widehat{\text{Corr}}(\hat{x},x) \, \sigma_{x} - \sigma_{\hat{x}}~ \bigg\}
    \\
    &= 
    \begin{dcases}
        ~(+) & \text{if } \beta_1 > 0 \text{ and } \widehat{\text{Corr}}(\hat{x},x) \, \sigma_{x} > \sigma_{\hat{x}} \\
        ~(-) & \text{if } \beta_1 > 0 \text{ and } \widehat{\text{Corr}}(\hat{x},x) \, \sigma_{x} < \sigma_{\hat{x}} \\
        ~(-) & \text{if } \beta_1 < 0 \text{ and } \widehat{\text{Corr}}(\hat{x},x) \, \sigma_{x} > \sigma_{\hat{x}} \\
        ~(+) & \text{if } \beta_1 < 0 \text{ and } \widehat{\text{Corr}}(\hat{x},x) \, \sigma_{x} < \sigma_{\hat{x}} \\
        ~\:\, 0 & \text{if } \beta_1 = 0 \text{ or } \widehat{\text{Corr}}(\hat{x},x) \, \sigma_{x} = \sigma_{\hat{x}}~,
    \end{dcases}
    \label{eq:sign-a}
\end{align}
where $\sigma_x$ refers to the standard deviation of $x$ ($\sigma_{\hat{x}}$ and $\sigma_{e}$ are defined similarly).

As (\ref{eq:sign-a}) reveals, the sign of $\beta_1 \widehat{\text{Cov}}(\hat{x},e)$ depends on two quantities: (\textit{i}) the sign of $\beta_1$, and (\textit{ii}) the sign of $\widehat{\text{Corr}}(\hat{x},x) \, \sigma_{x} - \sigma_{\hat{x}}$. It is difficult to generalize the sign of $\widehat{\text{Cov}}(\hat{x},e)$ without further assumptions. While one may be tempted to assume $\sigma_x > \sigma_{\hat{x}}$, this assumption is not sufficient for signing $\widehat{\text{Cov}}(\hat{x},e)$, as it still depends upon the magnitude of $\widehat{\text{Corr}}(\hat{x},x)$.\footnote{Further, this assumption is equivalent to making an assumption on $\widehat{\text{Cov}}(\hat{x},e)$, which means one is essentially assuming the result. That is, $\widehat{\text{Var}}(x) = \widehat{\text{Var}}(\hat{x}) + \widehat{\text{Var}}(e) + 2~\widehat{\text{Cov}}(\hat{x}, e)$. That said, in every iteration of our simulations, $\widehat{\text{Var}}(x) > \widehat{\text{Var}}(\hat{x})$.} The knife-edge case where $a=0$ appears unlikely except in cases where either $\beta_1=0$ or where $\widehat{\text{Cov}}(\hat{x},e)$ is mechanically zero (\eg OLS).

Across the twelve models that we consider in Table~\ref{tab:bias-decomp}, only the non-OLS models produce $\widehat{\text{Cov}}(\hat{x},e)\neq 0$---this is unsurprising. Lasso, neural nets, boosted trees, and random forests all produce positive covariance between $\hat{x}$ and $e$. In other words, in all of our DGPs, the term $\widehat{\text{Cov}}(\hat{x},e)$ biases $\hat{\beta}$ upward (positively) whenever it is non-zero.\footnote{This upward bias is partly due to the true parameter $\beta_1$ being positive.} Random forest models generate the largest covariance between $\hat{x}$ and $e$ (and consequently the largest $\widehat{\text{Cov}}(\hat{x},e)$) in each of the DGPs. Depending upon the DGP, Lasso, neural nets, and boosted trees generate the second-highest covariance. Because our \textit{shallow} subcase of neural nets approximates OLS, its covariance between $\hat{x}$ and $e$ is approximately zero.

One way to ensure that $\widehat{\text{Cov}}(\hat{x},e)=0$ for a nonlinear model is to linearize its output---this is achievable, for example, by using the ML-based prediction $\hat{x}(\mathbf{z})$ as an \textit{instrument} for $x$, rather than plugging it into the second stage \citep{Angrist2001,Chen2020-IV}. While this approach forces $\widehat{\text{Cov}}(\hat{x},e)=0$, it requires strengthening assumptions. We discuss this possibility further in Section~\ref{sec:discussion}.

More broadly, the component of bias due to covariance between first-stage predictions $(\hat{x})$ and their residuals $(e)$ (\ie the $\widehat{\text{Cov}}(\hat{x},e)$ term) accounts for the vast majority of the bias for Lasso and substantial amounts of the bias in random forests, boosted trees, and neural nets (the exact portion of the bias differs across DGPs and iterations). While $\widehat{\text{Cov}}(\hat{x},e)$ does not account for all of the bias, the non-zero covariance between first-stage predictions and residuals is an important (potentially large) component of the bias of ML-based 2SLS models.

\subsubsection*{The $\widehat{\mathrm{Cov}}(\hat{x},u)$ component}

Unlike $\widehat{\text{Cov}}(\hat{x},e)$, the second component of the wedge between $\hat\beta_1^{\text{2SLS}}$ and $\beta_1$ can be non-zero for both OLS-based methods and non-OLS models.
However, methods that use \textit{non-linear} predictions of $x$ in the first stage (\ie ML-assisted 2SLS) require special care to reduce $\widehat{\text{Cov}}(\hat{x},u)$ and produce low-bias estimates of $\beta_1$. 

This second term, $\widehat{\text{Cov}}(\hat{x},u)$, is effectively the exclusion restriction, and any 2SLS-inspired estimator can reduce bias in $\hat{\beta_1}$ by ensuring $\widehat{\text{Cov}}(\hat{x},u)$ is approximately zero. Assuming the instruments $\mathbf{z}$ are valid, an arbitrary prediction algorithm can maintain $\widehat{\text{Cov}}(\hat{x},u)\approx 0$ through either of three conditions:
\begin{enumerate}
    \item \textbf{Restrict the algorithm's choice set:} By restricting the learning algorithm to choosing from a set/class of functions where each individual function satisfies the exclusion restriction, one mechanically ensures the first-stage predictions $\hat{x}$ do not co-vary with the unobserved disturbance $u$. For example, traditional OLS-based 2SLS chooses the first stage from the set of linear combinations of the instruments. As long as the instruments are uncorrelated with the disturbance, any element of this set (the linear combination of the instruments) will also be uncorrelated with the disturbance---satisfying the exclusion restriction.
    \item \textbf{Extend the exclusion restriction:} One may extend the commonly invoked ``uncorrelated'' assumption to the stronger and underlying exclusion restriction of conditional mean independence. Rather than only assuming all \textit{linear} combinations of the instruments are exogenous, one could assume that \textbf{all functions} of the instruments (linear and nonlinear) satisfy the exclusion restriction. Put differently, this condition requires $\widehat{\text{Cov}}(f(\mathbf{z}),u) \approx 0$ for any function $f$. 
    \item \textbf{Really \textit{trust} the ML algorithm:} The final option is to simply rely upon the algorithm to find a function that satisfies the exclusion restriction, irrespective of choice set---something akin to closing one's eyes and hoping for the best. While this option makes a heroic assumption, as ML algorithms are not typically designed to search for and find valid exclusion restrictions, it is the default scenario. If a practitioner does not enforce condition {\bf 1} and does not assume condition {\bf 2}, then she is left hoping that the ML methods successfully chose a function that includes a valid exclusion restriction.
\end{enumerate}


Put simply, sufficiently flexible learning algorithms can recover endogenous variation in $x$---even when \textit{using linearly valid instruments}. Many ML training methods explicitly incentivize and enable algorithms to do this. 

Column $b$ of Table~\ref{tab:bias-decomp} highlights the tendency of flexible first-stage models (\eg tree methods and neural nets) to recover endogeneity. As the flexibility of algorithms increases, $\widehat{\text{Cov}}(\hat{x},u)$ tends to increase as well (across all DGPs). This covariance and its associated bias are particularly large for tree-based methods (especially random forests) and neural nets with multiple hidden layers. Notably, in panels~\ref{fig:density-coef-t1}--\ref{fig:density-coef-t3} of Figure~\ref{fig:results-densities}, the densities of unrestricted and narrow neural networks are bimodal. As Appendix Figure~\ref{fig:nn-depth} illustrates, the bimodality results from whether the neural network (\textit{i}) ``chooses'' zero hidden layers (the less biased mode) or (\textit{ii}) goes deeper (learning the endogenous error and generating more bias).\footnote{This result highlights the importance of allowing neural networks to choose no hidden layers.}\footnote{Another related concern familiar to the ML literature is \textit{overfit}. Overfit models tend to produce larger values of $\widehat{\text{Cov}}(\hat{x},u)$ than models that have been cross-validated. Though cross-validation is best/standard practice for machine-learning methods in prediction problems, here it retains importance by preventing the algorithms from overfitting the target variable $x$ in the first stage (even when out-of-sample performance is no longer the goal). We use five-fold cross-validation (CV) to tune the hyperparameters for Lasso-, tree-, and neural-net-based methods. Our neural-net cross-validation departs from standard five-fold CV. In Appendix Section~\ref{sec:MLP} we detail our cross-validation process for training neural net. One might further avoid overfit by applying holdout-style methods---only generating predictions for observation $i$ when $i$ is not in the training set. JIVE, split-sample IV, and \cite{Chen2020-IV} all feature this additional safeguard. We do not employ these holdout-based methods because our goal in this paper is to simulate the results of a researcher using off-the-shelf ML tools in the first stage of 2SLS.} This covariance between predictions $\hat{x}$ and the unobserved disturbance $u$ accounts for a substantial amount of the bias in nonlinear methods, which demonstrates that the previously discussed first component $\widehat{\text{Cov}}(\hat{x},e)$ is not the only issue facing these models.

\subsubsection*{The $\dfrac{1}{\mathrm{Var}(\hat{x})}$ component}

While the first two bias components enter additively, the third component scales their sum. Any method that reduces the variance of the first-stage predictions (\ie reduces $\widehat{\text{Var}}(\hat{x})$) mechanically inflates the bias produced by $\beta_1\widehat{\text{Cov}}(\hat{x},e) + \widehat{\text{Cov}}(\hat{x},u)$.

In the case of properly specified, OLS-based 2SLS, the variance of the predictions hardly affects bias in $\beta_1$, since $\widehat{\text{Cov}}(\hat{x},e)=0$ and $\widehat{\text{Cov}}(\hat{x},u)\approx 0$. However, most ML algorithms reduce the variance of their predictions while trading between out-of-sample bias and variance. This tradeoff between bias and variance happens {\it outside of a 2SLS framework}. Consequently, when practitioners insert variance-reducing ML methods into 2SLS, the variance reduction actually amplifies bias in the second-stage estimates.

Taking these insights to the results in Panel B of  Table~\ref{tab:bias-decomp}, notice that variance reduction can cause methods to perform poorly. For example, Lasso-assisted 2SLS produces the lowest variance $\hat{x}$ in two of the three high-complexity cases. (In cases 2 and 3, Lasso-based 2SLS has the highest $1/\widehat{\text{Var}}{\hat{x}}$-based amplifier of the bias.) This high degree of bias amplification generates notable bias in Lasso relative to many other methods. (This is also evident in Figure~\ref{fig:results-densities}). So while Lasso's $\widehat{\text{Cov}}(\hat{x},e)$ and $\widehat{\text{Cov}}(\hat{x},u)$ are less than or equal to those of many other methods, the amplification produced by variance-reduction in $\hat{x}$ ultimately causes Lasso to have substantial bias. Notably, post-Lasso-based 2SLS produces less bias, partly due to the fact that it includes less variance reduction.

Worse yet, tree-based methods substantially reduce variance \textit{and} produce relatively large $\widehat{\text{Cov}}(\hat{x},e)$ and $\widehat{\text{Cov}}(\hat{x},u)$ components, resulting in very large bias in their parameter estimates (even larger than na\"ive OLS).


\section{Discussion} \label{sec:discussion}

In this paper, we examine the implications of plugging off-the-shelf ML methods into a 2SLS framework. In many cases, injecting ML into the first stage of 2SLS generates substantial bias.

While there are many approaches to combining instrumental-variable intuition and machine learning, they relax the traditional 2SLS structure and require different (generally stronger) identifying assumptions and/or tailor-made ML algorithms.\footnote{For example, MLSS \citep{Chen2020-IV}, DeepIV \citep{Hartford2017}, DeepGMM \citep{Bennett2020}, KIV \citep{Singh2019}, Adversarial Estimation of Riesz Representers \citep{chernozhukov2020}, Neural Estimation of SEM \citep{liao2020}, and Non-Parametric IV \citep{kilbertus2020class}.} Among the current options, the closest in spirit to our question of ``What are the implications of inserting ML into 2SLS?" is the ``machine learning split-sample" (MLSS) estimator proposed by \cite{Chen2020-IV}.\footnote{\cite{Angrist2020} also applies split-sample methods to several ML algorithms (\ie post-Lasso, random forest), both in the first stage of 2SLS and while synthesizing instruments in a stage that precedes the first stage.}

With two fairly simple expansions of the traditional 2SLS framework, MLSS mitigates many biases generated by na\"ively plugging ML methods into the first stage. However, as with other more ML-forward methods, the solution is not without the cost of substantially strengthening the exclusion restriction. Specifically, \cite{Chen2020-IV} proposes augmenting 2SLS with two simple techniques: restrict ML-based predictions to be explicitly out of sample (using split-sample methods) and use the ML-generated predictions as a ``synthetic'' instrument that then enters linearly in the first stage. 

The idea for out-of-sample (split-sample) ML predictions follows the lead of Jackknife IV and Split Sample IV. By introducing out-of-sample methods to the ML-prediction exercise, \citeauthor{Chen2020-IV} aims to prevent the ML algorithm from fitting the first-stage errors, and shutting down the bias generated by $\widehat{\text{Cov}}(\hat{x},u)$. One potential drawback, however, is that this out-of-sample step likely increases variability (as seen in the JIVE results of Figure~\ref{fig:results-densities}).

The second component of \citeauthor{Chen2020-IV} involves a \textit{zero\textsuperscript{th} stage} (i.e., before the first stage), in which the practitioner trains an ML algorithm to predict $x$ using the instruments $\mathbf{z}$.\footnote{Note that this zero\textsuperscript{th} stage is identical to first stages that na\"ively insert ML methods into 2SLS.} The predictions from this zero\textsuperscript{th} stage are then used as the instrument within a traditional 2SLS framework. The benefit is that the resulting linear first stage (linearizing the results of a potentially forbidden regression) guarantees that $\widehat{\text{Cov}}(\hat{x},e)=0$ and shuts down one avenue through which bias enters. 

Importantly, this zero\textsuperscript{th} stage of the MLSS approach requires that no learnable function of instruments meaningfully predicts the structural disturbance $u$. The function spaces of ML algorithms can cover all possible functions of the instruments (\eg most neural networks are universal approximators), which requires strengthening the exclusion restriction from the assumption of ``no correlation'' to the actual underlying assumption of conditional mean-independence between the instruments and disturbance. For example, this strengthening includes ML-learned step functions, threshold indicators, kinks, many-way interactions---\textit{any} learnable function $f(z)$ that is informative for $x$. This potentially infinite-dimensional set of exclusion restrictions is presumably more difficult to justify than the typical identifying assumption assumed in 2SLS applications.

Overall, solutions exist for the bias components that we identify in this paper, but each solution comes at a cost. Some solutions are fairly cheap (\eg out-of-sample prediction methods require sufficient overlap and can increase noise). Other solutions require more from the practitioner, \eg strengthening a linear exclusion restriction to an infinite-dimensional function space. Researchers may be comfortable invoking these stronger identifying assumptions in some settings. However, more work is needed in exploring and bounding the relative benefits and costs of these procedures in practice.\footnote{For example, future work could continue to explore the bias from instruments that nonlinearly violate an expanded exclusion restriction (potentially discoverable by nonlinear ML algorithms) but satisfy the traditional linear exclusion restriction. Future work could also consider how ML-based 2SLS approaches estimate potentially different LATEs.}

\section{Conclusion} \label{sec:conclussion}


Our results show that na\"ively inserting machine-learning algorithms into the first stage of a 2SLS structure often produces bias. While some channels of this bias will be familiar to an economics audience, others may be less familiar, and arise from the intersection of ``classical" econometric tools with ML methods. In terms of minimum bias, we find that the most-successful cases of integrating machine learning into 2SLS are largely restricted to instrument selection or modification (\eg post-Lasso) before entering a traditional 2SLS framework. However, inserting highly nonlinear algorithms (\eg random forests, boosted trees, and relatively deep neural nets) into the first stage of 2SLS drives estimates of causal parameters to be {\it more} biased than much simpler and more-easily interpreted alternatives. Importantly, such methods can even yield more bias than a na\"ive, endogenous OLS regression.

We mechanically decompose the bias within ML-augmented 2SLS estimates into three terms: 
\begin{itemize}
    \item $\beta ~\widehat{\text{Cov}}(\hat{x},e)$: When first-stage predictions are not orthogonal to their residuals (relevant for nonlinear methods), $\widehat{\text{Cov}}(\hat{x},e)\neq0$. This wedge increases with the magnitude of the target parameter, $\beta$. 
    \item $\widehat{\text{Cov}}(\hat{x},u)$: Flexible ML methods can overfit $x$ in the first stage, unintentionally recovering endogenous variation in $x$. This \textit{overfit} first stage produces fitted values $(\hat{x})$ that are no longer exogenous from the structural disturbance $u$, even when the first-stage instruments are exogenous.
    \item $1/\widehat{\text{Var}}(\hat{x})$: ML-based prediction methods typically reduce the variance of their predictions. As such, ML-augmented 2SLS often reduces the variance of $\hat{x}$, which amplifies any non-zero wedge between the estimate and the estimand coming from $\beta ~ \widehat{\text{Cov}}(\hat{x},e) + \widehat{\text{Cov}}(\hat{x},u)$.
\end{itemize}




Overall, we find that many of the methods pioneered in the 1990s and 2000s (\eg split-sample IV, LIML, and JIVE) reduce bias resulting from weak or over-identified IV for a linear first stage.\footnote{\cite{Angrist2020} comes to a similar conclusion. Focusing more on heterogeneous treatment effects and ML-assisted covariate selection, \citeauthor{Angrist2020} does not decompose the bias associated with ML methods applied to the instruments in the first-stage prediction problem (Section~\ref{ssec:decomp} and summarized above). Others have also found that JIVE and LIML are the best choice in the weak-many-instruments case---\cite{Hansen2014} explores a regularized JIVE, and \cite{Carrasco2015} develops a regularized LIML approach.} At the same time, we find that less-traditional methods that \textit{mindfully} incorporate machine learning techniques (\eg PCA-synthesized instruments and post-Lasso) generate improvements over 2SLS across a variety of DGPs.\footnote{See \cite{Ackerberg2009} for a discussion of tools for addressing over-identified instrumental variables estimation and \cite{Andrews2019} for a discussion of weak instruments in linear IV regression.} Importantly, we show that carelessly injecting ML into the first stage can also produce substantial bias, possibly worsening bias over na\"ive OLS.

The fundamental problem comes from recognizing the first stage of 2SLS solely as a prediction problem without recognizing that it is also part of a larger estimation {\it system}.

ML methods can produce good predictions. However, they were neither developed nor optimized for two-stage procedures that generate minimally biased causal estimates. For example, variance reduction in prediction typically improves out-of-sample prediction accuracy but can actually inflate parameter bias in the second stage of 2SLS. Solutions have been proposed to these and other issues from ML-based 2SLS. However, if one approaches these problems without the caution that we suggest (\eg blindly relying upon ML to learn the variation in the first stage) bias often results. Unfortunately, the solutions to these ML-in-2SLS problems can often require the practitioner to strengthen the exclusion restriction where complexity is highest (\eg in high-dimensional data). Because ML methods are typically adopted in complex data environments that are not well understood, employing these methods without safeguards can entice researchers into making heroic assumptions. 

Given the performance of existing off-the-shelf estimators, na\"ively injecting ML into 2SLS appears to produce little gain relative to its costs/biases. More broadly, applying ML methods to 2SLS requires the practitioner to face issues present in both the prediction \textit{and} causal-inference settings---and in addition, less-familiar issues arising from their interaction. 

\newgeometry{margin=1in}
\clearpage
\section*{Figures}

\begin{landscape}
\begin{figure}
    \caption{\textbf{Exclusion-restriction violations} via higher-order interactions among instruments (`low-complexity case' of 7 strong instruments)}
    \label{fig:interactions}
    \centering
    \begin{subfigure}{0.5\linewidth}
        \centering
        \caption{\textbf{`Standard' linear estimators:} OLS, LIML, SSIV, and JIVE}
        \label{fig:interactions-linear}
        \includegraphics[width=0.99\linewidth]{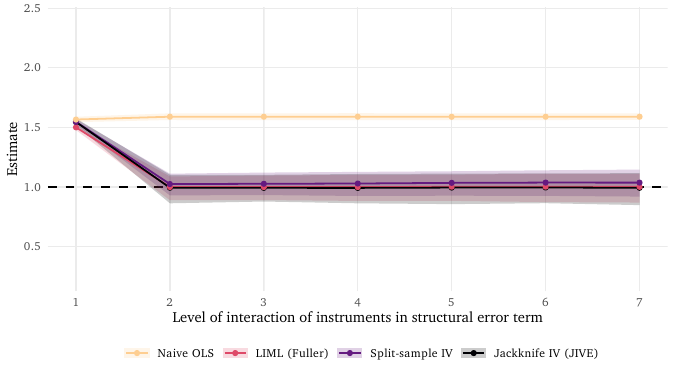}
    \end{subfigure}%
    \begin{subfigure}{0.5\linewidth}
        \centering
        \caption{\textbf{`Selection' methods:} Lasso, post-Lasso, and PCA}
        \label{fig:interactions-selection}
        \includegraphics[width=0.99\linewidth]{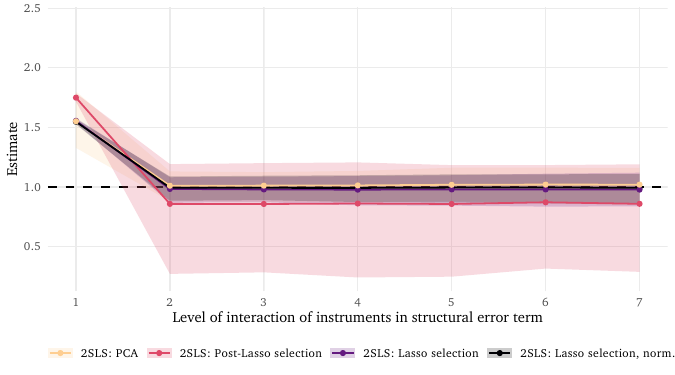}
    \end{subfigure}%
    \hfill
    \begin{subfigure}{0.5\linewidth}
        \centering
        \caption{\textbf{Trees:} Random forests and boosted trees---with and without normalization}
        \label{fig:interactions-trees}
        \includegraphics[width=0.99\linewidth]{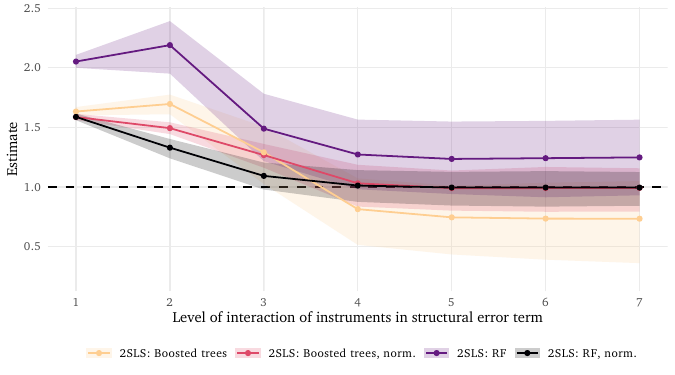}
    \end{subfigure}%
    \begin{subfigure}{0.5\linewidth}
        \centering
        \caption{\textbf{Neural networks:} With and without normalization}
        \label{fig:interactions-nets}
        \includegraphics[width=0.99\linewidth]{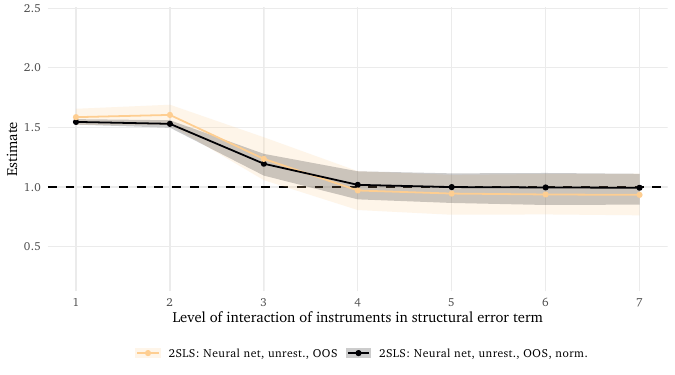}
    \end{subfigure}%
    \caption*{\small The DGPs underlying these figures add a $k$-term interaction between the instruments to the structural error. \textit{E.g.}, when the $x$-axis equals $3$, we add the interaction $x_1 \times x_2 \times x_3$ to the error; $k=1$ is a linear exclusion-restriction violation. Interactions with $k>1$ do not violate the exclusion restriction of linear methods but do violate the exclusion restriction for methods that require a higher-order/expanded exclusion restriction (\ie conditional independence). Each subfigure illustrates the performance of the given estimators from 1,000 iterations. The solid lines and dots mark the mean of the estimator-DGP combination; the shaded bands give the boundaries between the 2.5th and 97.5th percentiles. The true value is 1. \textit{Normalized} methods (`norm.') use the stated method to synthesize a single instrument from the seven instruments \textit{prior to the first stage} (as in \cite{Chen2020-IV}). See  Figure~\ref{fig:interaction-densities} for the densities of 2-, 3-, 4-, and 5-way interactions.}
\end{figure}
\end{landscape}

\clearpage
\section{Tables}

 
 \begin{table}[!htbp] \centering
   \caption{\textbf{Simulation results}: Mean and standard deviation for methods and DGPs}
   \label{tab:main-results}
 \begin{tabular}{@{\extracolsep{4pt}}lcccc@{}}
   \toprule
   \ra{1.5}
  & \textit{Low-complexity} case & \multicolumn{3}{c}{\textit{High-complexity} cases} \\ \cmidrule(lr){2-2} \cmidrule(lr){3-5}
   & 7 strong instruments & \multicolumn{3}{c}{100 \textit{mixed} instruments} \\ 
   & ($A$) & ($B$) & ($C$) & ($D$) \\
   \midrule
 
 Naive OLS                           &    1.038     &    1.335     &    1.334     &    1.223     \\[-0.1em] 
                                     &    (0.007)   &    (0.020)   &    (0.019)   &    (0.046)   \\[0.3em]  
 First stage: OLS                    &    1.000     &    1.058     &    1.056     &    1.023     \\[-0.1em] 
                                     &    (0.007)   &    (0.040)   &    (0.039)   &    (0.042)   \\[0.3em]  
 LIML (Fuller)                       &    1.000     &    1.000     &    0.998     &    1.000     \\[-0.1em] 
                                     &    (0.008)   &    (0.044)   &    (0.043)   &    (0.043)   \\[0.3em]  
 Split-sample IV                     &    1.001     &    1.062     &    1.060     &    1.025     \\[-0.1em] 
                                     &    (0.007)   &    (0.041)   &    (0.040)   &    (0.043)   \\[0.3em]  
 Jackknife IV (JIVE)                 &    1.000     &    0.998     &    0.996     &    1.000     \\[-0.1em] 
                                     &    (0.017)   &    (0.044)   &    (0.044)   &    (0.044)   \\[0.3em]  
 First stage: PCA                    &    1.000     &    1.032     &    1.026     &    1.016     \\[-0.1em] 
                                     &    (0.007)   &    (0.042)   &    (0.041)   &    (0.045)   \\[0.3em]  
 First stage: Post-Lasso selection   &    1.000     &    1.026     &    1.023     &    1.013     \\[-0.1em] 
                                     &    (0.007)   &    (0.044)   &    (0.042)   &    (0.042)   \\[0.3em]  
 First stage: Lasso                  &    1.007     &    1.100     &    1.098     &    1.042     \\[-0.1em] 
                                     &    (0.007)   &    (0.045)   &    (0.045)   &    (0.045)   \\[0.3em]  
 First stage: Neural net             &    1.008     &    1.215     &    1.209     &    1.110     \\[-0.1em] 
                                     &    (0.029)   &    (0.180)   &    (0.176)   &    (0.105)   \\[0.3em]  
 First stage: Neural net, shallow    &    1.002     &    1.069     &    1.065     &    1.030     \\[-0.1em] 
                                     &    (0.018)   &    (0.066)   &    (0.067)   &    (0.049)   \\[0.3em]  
 First stage: Neural net, narrow     &    1.008     &    1.213     &    1.210     &    1.100     \\[-0.1em] 
                                     &    (0.027)   &    (0.183)   &    (0.185)   &    (0.103)   \\[0.3em]  
 First stage: Boosted trees          &    1.008     &    1.254     &    1.255     &    1.121     \\[-0.1em] 
                                     &    (0.007)   &    (0.041)   &    (0.039)   &    (0.047)   \\[0.3em]  
 First stage: Random forest, CV      &    1.071     &    1.562     &    1.563     &    1.316     \\[-0.1em] 
                                     &    (0.008)   &    (0.033)   &    (0.034)   &    (0.058)   \\[0.3em]  
 
 \bottomrule
 \\[-0.9em]
 \end{tabular}
 \caption*{\raggedright\small This table provides the means and standard deviations of the distributions illustrated in Figure~\ref{fig:results-densities}. Each \textbf{column} contains a separate DGP: (a) contains the \textit{low-complexity} DGP with 7 (equally) strong instruments; (b)--(d) contain the three \textit{high-complexity} cases with 100 instruments of mixed strengths. \textbf{Rows} differ by estimator. For each DGP-estimator combination, we summarize the estimates for the parameter of interest $(\beta)$ across 1,000 iterations using a mean and standard deviation (the standard deviation is in parentheses). }
 \end{table}

 
 \begin{table}[!htbp] \centering
   \caption{\textbf{Simulation results}: Decomposing bias components (means from simulation)}
   \label{tab:bias-decomp}
 \resizebox{0.9\columnwidth}{!}{%
 \begin{tabular}{@{\extracolsep{-5pt}}lccccccccc@{}}
   \toprule
   \ra{1.5}
 
  & & \multicolumn{3}{c}{\textbf{Bias components}} \\ \cmidrule(lr){3-5}
  & $(a+b)c$ & $a$ & $b$ & $c$ \\
 \enspace \textbf{Estimator} & Bias & $\widehat{\text{Cov}}(\hat{x},e)$ & $\widehat{\text{Cov}}(\hat{x},u)$ & $1/\widehat{\text{Var}}(\hat{x})$ & $\widehat{\text{Var}}(\hat{x})$ & $\widehat{\text{Var}}(x)$ & $\widehat{\text{Cov}}(x,\hat{x})$ & $\text{Corr}(x,\hat{x})$ & $\widehat{\text{Cov}}(x,u)$ \\
 \midrule
 \multicolumn{3}{@{}l}{\textbf{(A)} \textit{Low-complexity} case} \\
 \enspace Naive OLS                           &    0.04   &    0      &    1.01   &    0.04   &    26.41   &    26.41   &    26.41   &    1      &    1.01   \\ 
 \enspace First stage: OLS                    &    0      &    0      &    0.01   &    0.04   &    25.41   &    26.41   &    25.41   &    0.98   &    1.01   \\ 
 \enspace Split-sample IV                     &    0      &    0      &    0.02   &    0.04   &    25.42   &    26.41   &    25.42   &    0.98   &    1.01   \\ 
 \enspace Jackknife IV (JIVE)                 &    0      &    0      &    0.01   &    0.05   &    20.76   &    21.74   &    20.76   &    0.98   &    1.01   \\ 
 \enspace First stage: PCA                    &    0      &    0      &    0.01   &    0.04   &    25.41   &    26.41   &    25.41   &    0.98   &    1.01   \\ 
 \enspace First stage: Post-Lasso selection   &    0      &    0      &    0.01   &    0.04   &    25.41   &    26.41   &    25.41   &    0.98   &    1.01   \\ 
 \enspace First stage: Lasso selection        &    0.01   &    0.17   &    0.01   &    0.04   &    25.08   &    26.41   &    25.25   &    0.98   &    1.01   \\ 
 \enspace First stage: Neural net             &    0.01   &    0.1    &    0.05   &    0.05   &    20.52   &    21.67   &    20.61   &    0.98   &    0.99   \\ 
 \enspace First stage: Neural net, narrow     &    0.01   &    0.09   &    0.04   &    0.05   &    20.52   &    21.67   &    20.61   &    0.98   &    0.99   \\ 
 \enspace First stage: Neural net, shallow    &    0      &    0      &    0.03   &    0.05   &    20.71   &    21.67   &    20.71   &    0.98   &    0.99   \\ 
 \enspace First stage: Boosted trees          &    0.01   &    0.01   &    0.2    &    0.04   &    25.51   &    26.41   &    25.52   &    0.98   &    1.01   \\ 
 \enspace First stage: Random forest, CV      &    0.07   &    1.08   &    0.62   &    0.04   &    24.01   &    26.41   &    25.09   &    1      &    1.01   \\ 
 \midrule
 \multicolumn{3}{@{}l}{\textbf{(B)} \textit{High-complexity} case 1} \\
 \enspace Naive OLS                           &    0.22   &    0      &    0.15   &    1.56   &    0.64   &    0.64   &    0.64   &    1      &    0.15   \\ 
 \enspace First stage: OLS                    &    0.02   &    0      &    0.01   &    1.77   &    0.56   &    0.64   &    0.56   &    0.94   &    0.15   \\ 
 \enspace Split-sample IV                     &    0.03   &    0      &    0.01   &    1.79   &    0.56   &    0.64   &    0.56   &    0.93   &    0.15   \\ 
 \enspace Jackknife IV (JIVE)                 &    0      &    0      &    0      &    1.77   &    0.57   &    0.65   &    0.57   &    0.94   &    0.15   \\ 
 \enspace First stage: PCA                    &    0.02   &    0      &    0.01   &    2.02   &    0.5    &    0.64   &    0.5    &    0.88   &    0.15   \\ 
 \enspace First stage: Post-Lasso selection   &    0.01   &    0      &    0.01   &    1.79   &    0.56   &    0.64   &    0.56   &    0.93   &    0.15   \\ 
 \enspace First stage: Lasso selection        &    0.04   &    0.02   &    0      &    1.92   &    0.52   &    0.64   &    0.54   &    0.93   &    0.15   \\ 
 \enspace First stage: Neural net, shallow    &    0.03   &    0      &    0.02   &    1.77   &    0.57   &    0.64   &    0.57   &    0.94   &    0.15   \\ 
 \enspace First stage: Neural net, narrow     &    0.1    &    0.01   &    0.05   &    1.75   &    0.57   &    0.64   &    0.58   &    0.96   &    0.15   \\ 
 \enspace First stage: Neural net             &    0.11   &    0.01   &    0.06   &    1.74   &    0.58   &    0.64   &    0.58   &    0.96   &    0.15   \\ 
 \enspace First stage: Boosted trees          &    0.12   &    0.03   &    0.04   &    1.91   &    0.52   &    0.65   &    0.55   &    0.95   &    0.15   \\ 
 \enspace First stage: Random forest, CV      &    0.32   &    0.06   &    0.09   &    2.04   &    0.49   &    0.65   &    0.56   &    0.99   &    0.15   \\ 
 \midrule
 \multicolumn{3}{@{}l}{\textbf{(C)} \textit{High-complexity} case 2} \\
 \enspace Naive OLS                           &    0.33   &    0      &    0.35   &    0.95   &    1.06   &    1.06   &    1.06   &    1      &    0.35   \\ 
 \enspace First stage: OLS                    &    0.06   &    0      &    0.03   &    1.66   &    0.6    &    1.06   &    0.6    &    0.76   &    0.35   \\ 
 \enspace Split-sample IV                     &    0.06   &    0      &    0.03   &    1.76   &    0.57   &    1.06   &    0.57   &    0.73   &    0.35   \\ 
 \enspace Jackknife IV (JIVE)                 &    0      &    0      &    0      &    1.65   &    0.61   &    1.06   &    0.61   &    0.76   &    0.35   \\ 
 \enspace First stage: PCA                    &    0.03   &    0      &    0.02   &    1.75   &    0.57   &    1.06   &    0.57   &    0.74   &    0.35   \\ 
 \enspace First stage: Post-Lasso selection   &    0.02   &    0      &    0.01   &    1.75   &    0.57   &    1.06   &    0.57   &    0.74   &    0.35   \\ 
 \enspace First stage: Lasso selection        &    0.1    &    0.04   &    0.01   &    2.11   &    0.48   &    1.06   &    0.52   &    0.73   &    0.35   \\ 
 \enspace First stage: Neural net             &    0.21   &    0.03   &    0.13   &    1.49   &    0.7    &    1.06   &    0.73   &    0.84   &    0.35   \\ 
 \enspace First stage: Neural net, narrow     &    0.21   &    0.04   &    0.12   &    1.52   &    0.68   &    1.06   &    0.71   &    0.84   &    0.35   \\ 
\enspace First stage: Neural net, shallow    &    0.06   &    0      &    0.04   &    1.63   &    0.62   &    1.06   &    0.62   &    0.76   &    0.35   \\ 
 \enspace First stage: Boosted trees          &    0.25   &    0.07   &    0.07   &    1.83   &    0.55   &    1.06   &    0.62   &    0.81   &    0.36   \\ 
 \enspace First stage: Random forest, CV      &    0.56   &    0.16   &    0.22   &    1.51   &    0.66   &    1.06   &    0.82   &    0.98   &    0.35   \\ 
 \midrule
 \multicolumn{3}{@{}l}{\textbf{(D)} \textit{High-complexity} case 3} \\
 \enspace Naive OLS                           &    0.34   &    0      &    0.35   &    0.95   &    1.06   &    1.06   &    1.06   &    1      &    0.35   \\ 
 \enspace First stage: OLS                    &    0.06   &    0      &    0.04   &    1.66   &    0.61   &    1.06   &    0.61   &    0.76   &    0.35   \\ 
 \enspace Split-sample IV                     &    0.06   &    0      &    0.04   &    1.76   &    0.57   &    1.06   &    0.57   &    0.73   &    0.35   \\ 
 \enspace Jackknife IV (JIVE)                 &    0      &    0      &    0      &    1.64   &    0.61   &    1.06   &    0.61   &    0.76   &    0.36   \\ 
 \enspace First stage: PCA                    &    0.03   &    0      &    0.02   &    1.76   &    0.57   &    1.06   &    0.57   &    0.73   &    0.35   \\ 
 \enspace First stage: Post-Lasso selection   &    0.03   &    0      &    0.02   &    1.74   &    0.58   &    1.06   &    0.58   &    0.74   &    0.35   \\ 
 \enspace First stage: Lasso selection        &    0.1    &    0.04   &    0.01   &    2.1    &    0.48   &    1.06   &    0.52   &    0.73   &    0.35   \\ 
 \enspace First stage: Neural net             &    0.22   &    0.03   &    0.13   &    1.49   &    0.69   &    1.06   &    0.73   &    0.84   &    0.36   \\ 
\enspace First stage: Neural net, narrow     &    0.21   &    0.04   &    0.12   &    1.53   &    0.67   &    1.06   &    0.71   &    0.84   &    0.36   \\ 
 \enspace First stage: Neural net, shallow    &    0.07   &    0      &    0.05   &    1.63   &    0.62   &    1.06   &    0.62   &    0.76   &    0.36   \\ 
 \enspace First stage: Boosted trees          &    0.25   &    0.07   &    0.07   &    1.83   &    0.55   &    1.06   &    0.62   &    0.81   &    0.36   \\ 
 \enspace First stage: Random forest, CV      &    0.56   &    0.16   &    0.22   &    1.51   &    0.66   &    1.06   &    0.82   &    0.98   &    0.35   \\ 
 
 \bottomrule
 \\[-0.9em]
 \end{tabular}%
 }
 \caption*{\raggedright\small A cell's value provides the given statistic's mean across 1,000 iterations. The panels (A--D) denote the four separate DGPs; rows refer to estimators. We omit LIML as it is not a two-stage method and thus does not produce $\hat{x}$. }
 \end{table}

\clearpage
\bibliography{biblio}

\clearpage
\singlespacing
\begin{appendices}

\section{}
  \setcounter{table}{0}
  \renewcommand{\thetable}{A\arabic{table}}
  \setcounter{figure}{0}
  \renewcommand{\thefigure}{A\arabic{figure}}
  \setcounter{equation}{0}
  \renewcommand{\theequation}{A\arabic{equation}}

\subsection{Appendix: Figures} 

\begin{figure}
    \caption{\textbf{Predictions \textit{vs.} estimation:} Comparing cross-validated prediction performance with bias in random-forest-based 2SLS}
    \label{fig:cv-bias-rf}
    \centering
    \begin{subfigure}{0.99\linewidth}
        \caption{In- and out-of-sample MSE for predictions of $x$}
        \label{fig:cv-bias-rf-1}
        \centering
        \includegraphics[width=0.99\linewidth]{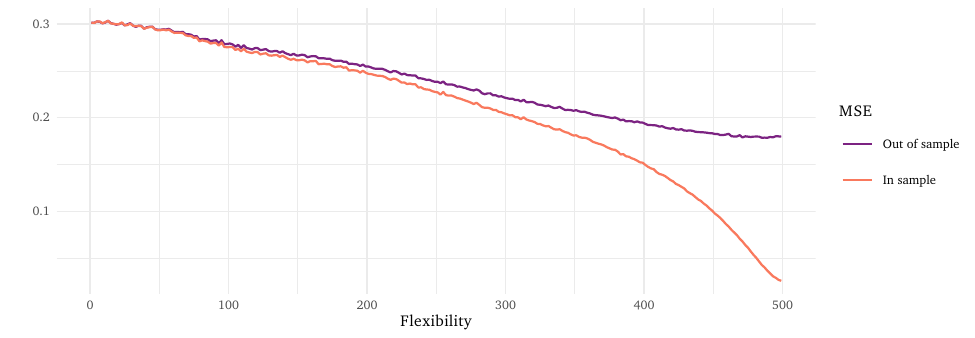}
    \end{subfigure}
    \begin{subfigure}{0.99\linewidth}
        \caption{The three `bias components' in estimating $\beta$}
        \label{fig:cv-bias-rf-2}
        \centering
        \includegraphics[width=0.99\linewidth]{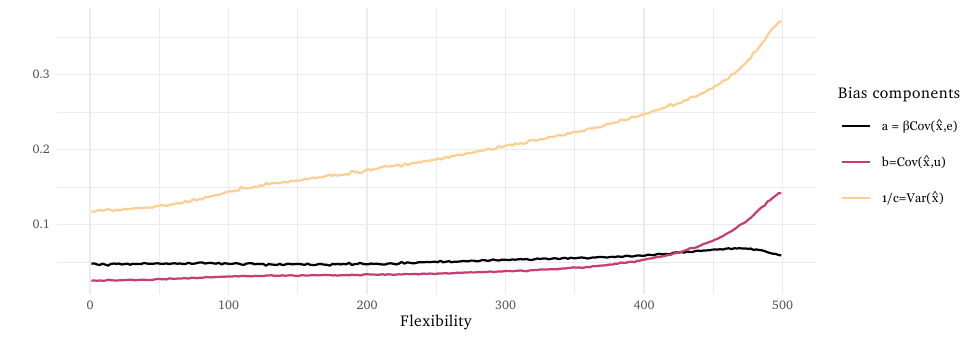}
    \end{subfigure}
    \begin{subfigure}{0.99\linewidth}
        \caption{Bias in $\hat{\beta}$ as a function of model flexibility ($\beta_1 = 1$)}
        \label{fig:cv-bias-rf-3}
        \centering
        \includegraphics[width=0.99\linewidth]{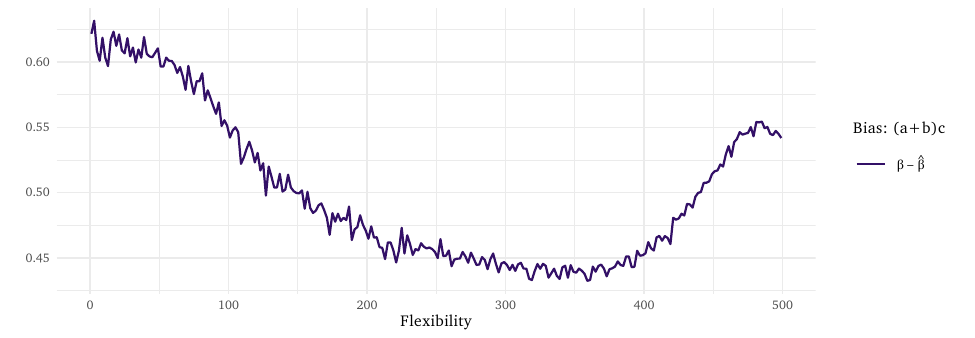}
    \end{subfigure}
    \hfill
\end{figure}

\begin{landscape}
\begin{figure}
    \caption{\textbf{Distributions of estimates} with ``exclusion-restriction violations'' from $k$-term interactions (`low-complexity case)}
    \label{fig:interaction-densities}
    \centering
    \begin{subfigure}{0.5\linewidth}
        \centering
        \caption{\textbf{Two-term `exclusion-restriction violation':} $x_1 \times x_2$}
        \label{fig:interaction-densities-2}
        \includegraphics[width=0.99\linewidth]{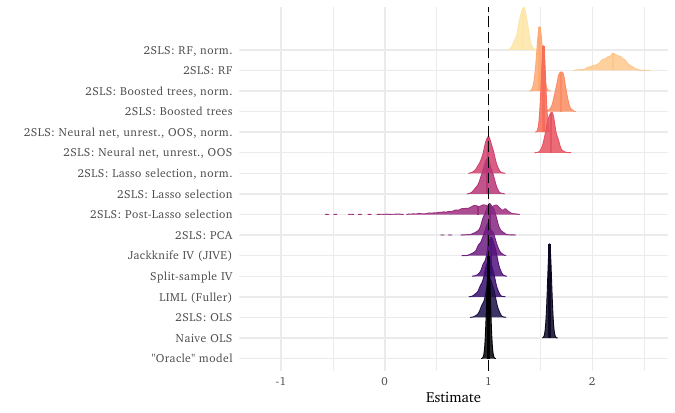}
    \end{subfigure}%
    \begin{subfigure}{0.5\linewidth}
        \centering
        \caption{\textbf{Three-term `exclusion-restriction violation':} $x_1 \times x_2 \times x_3$}
        \label{fig:interaction-densities-3}
        \includegraphics[width=0.99\linewidth]{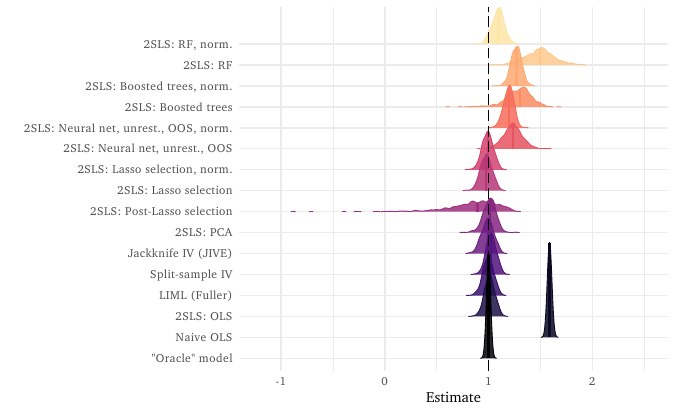}
    \end{subfigure}%
    \hfill
    \begin{subfigure}{0.5\linewidth}
        \centering
        \caption{\textbf{Four-term `exclusion-restriction violation':} $x_1 \times x_2 \times x_3 \times x_4$}
        \label{fig:interaction-densities-4}
        \includegraphics[width=0.99\linewidth]{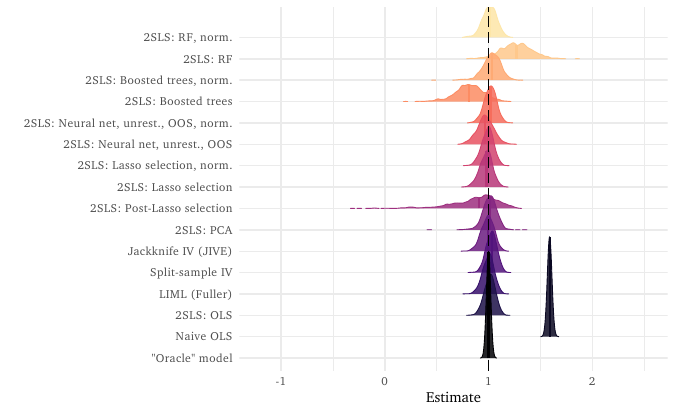}
    \end{subfigure}%
    \begin{subfigure}{0.5\linewidth}
        \centering
        \caption{\textbf{Five-term `exclusion-restriction violation':} $x_1 \times x_2 \times x_3 \times x_4 \times x_5$}
        \label{fig:interaction-densities-5}
        \includegraphics[width=0.99\linewidth]{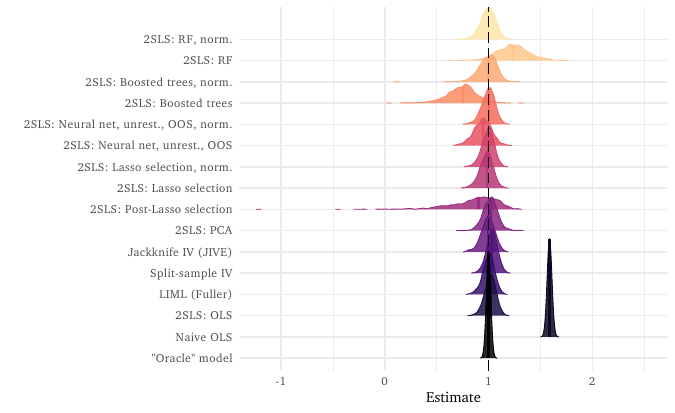}
    \end{subfigure}%
    \caption*{\footnotesize This figure illustrates the densities of the estimates portrayed in Figure~\ref{fig:interactions}.  Post-lasso's high-variance in this experiment come from using the so-called plug-in lambda value that prevents overselection of poor-performing instruments. To calculate this plug-in penalty, two parameters are user-chosen: $c$ and $\gamma$ and multiple theoretical values are used. We used the suggested values as detailed in \cite{Belloni2012} and \cite{Belloni2013}. This lambda plug-in value is calculated to be equal to $(1.1*2)\sqrt{N}*\Phi(\frac{1 - (\gamma)}{2*nz})$ and $\gamma \equiv \frac{.1}{ln(N \vee nz)}$. In our case, the number of instruments $nz = 7$, the number of observations $N = 1000$, and $\Phi \sim Quantile\ Normal$. This results in $\lambda \approx 36$. By using the suggested higher levels of $\lambda$, post-lasso can result in a higher chance of under-fit, with the extreme example being a second-stage regression model of the form f(X|Z) = 0. Given our variables are independent and relatively strong, regularization is highly unlikely to improve fit, and large lambda may result in under-specification for the first stage.}
\end{figure}
\end{landscape}

\clearpage
\begin{figure}
\caption{Explaining unrestricted/narrow neural networks' bimodal distributions of $\hat{\beta}$}
\label{fig:nn-depth}
\begin{minipage}[c][7.2in][t]{0.5\textwidth}
  \vspace*{\fill}
  \centering
  \subcaption{Comparing bias in $\hat{\beta}$: Approximately linear (no hidden layers) \textit{vs.} `deeper' neural networks}
  \label{fig:nn-depth-bias}
  \includegraphics[width=0.99\linewidth]{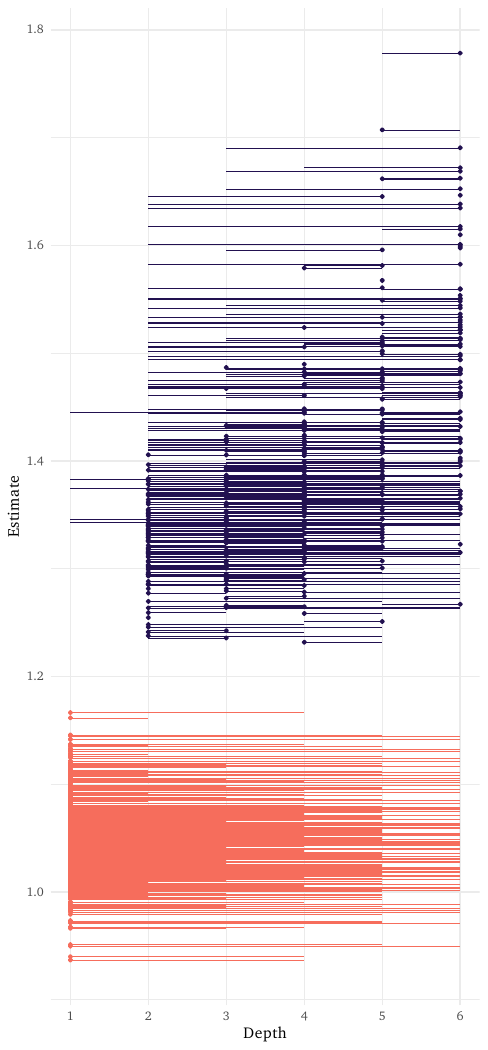}
\end{minipage}%
\begin{minipage}[c][7.2in][t]{0.5\textwidth}
  \vspace*{\fill}
  \centering
  \subcaption{Comparing bias in $\hat{\beta}$ and out-of-sample loss: Approximately linear \textit{vs.} `deeper' models}
  \label{fig:nn-depth-loss}
  \includegraphics[width=0.99\linewidth]{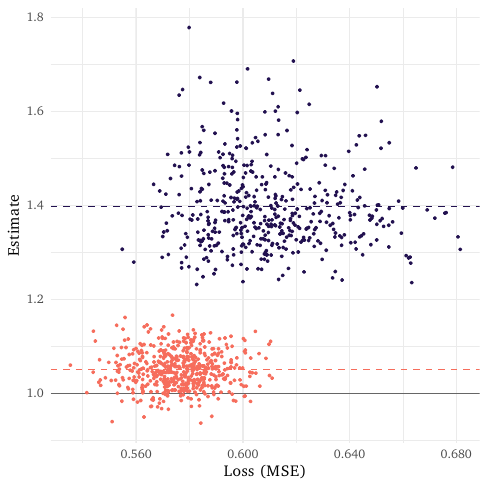}\par\vfill
  \subcaption{Cross validation's likelihood of choosing more shallow models when choosing between two options}
  \label{fig:nn-depth-choice}
  \includegraphics[width=0.99\linewidth]{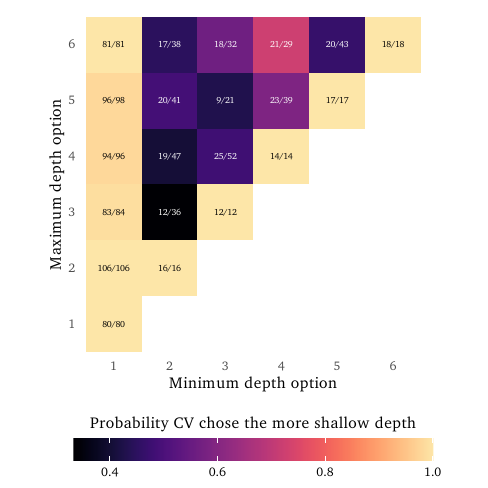}
\end{minipage}
\bigskip \caption*{\footnotesize The $y$ axis of Panel \textbf{a} depicts the second-stage estimate $\hat{\beta}$, and the $x$ axis represents the depths of the neural networks cross validated in each of the 1,000 iterations. ``Depth 1'' implies no hidden layers---directly linking the input and output (approximating linear regression). Horizontal line segments in \textbf{a} connect the two possible depths that the model chose between. The solid dot marks the chosen depth (by cross validation). In \textbf{a} and \textbf{b} we color the two subsets separately to illustrate the source of the bimodal distributions in Figure~\ref{fig:results-densities}. Panel \textbf{c} gives the probability that a model chose the more shallow (less deep) model for each combination of choices (probabilities are equal to the fractions in the cells). For instance, the top-left corner tells us that when facing a choice between a depth-1 model and a depth-6 model, CV chose the depth-1 model 81 out of 81 times. Each of these figures uses the results from 1,000 iterations of neural-net based 2SLS \textit{high-complexity case 2} (Panel C of Figure~\ref{fig:density-coef-t2}; 100 mixed-strength instruments with strength decreasing from $z_1$) with no restrictions on the hyperparameter space (\textit{unrestricted} in tables~\ref{tab:main-results}--\ref{tab:bias-decomp}). The results are very similar for the other high-complexity DGPs.}
\end{figure}


\subsection{Appendix: Math}

\subsubsection{Wedge \textit{a} covariance and correlation}
\label{sec:wedge-a}

Recall that $e = x - \hat{x}$, \ie $e$ is the first-stage prediction's residual.
\begingroup
\addtolength{\jot}{1em}
\begin{align}
    \nonumber
    \text{Corr}(\hat{x},e) 
    &= \dfrac{\text{Cov}(\hat{x},e)}{\sigma_{\hat{x}}\sigma_{e}}
    \\ \nonumber
    &= \dfrac{\text{Cov}(\hat{x},x)}{\sigma_{\hat{x}}\sigma_{e}} - \dfrac{\text{Var}(\hat{x})}{\sigma_{\hat{x}}\sigma_{e}}
    \\ \nonumber
    &= \dfrac{\text{Cov}(\hat{x},x)}{\sigma_{\hat{x}}\sigma_{e}} \dfrac{\sigma_{x}}{\sigma_{x}} - \dfrac{\sigma_{\hat{x}}}{\sigma_{e}}
    \\ \nonumber
    &= \dfrac{\text{Corr}(\hat{x},x) \, \sigma_{x}}{\sigma_{e}} - \dfrac{\sigma_{\hat{x}}}{\sigma_{e}}
    \\
    &= \sigma_{e}^{-1} \bigg( \text{Corr}(\hat{x},x) \, \sigma_{x} - \sigma_{\hat{x}} \bigg)
    \label{eq:corr-xhat-e}
\end{align}
\endgroup

\subsection{Appendix: MLP/Neural cross-validation procedure} \label{sec:MLP}

Unlike many of the other methods explored in this paper, MLP (Multi-Layer Perceptrons) are difficult to cross-validate in a consistent way. This is for three main reasons. 

The first reason is due to one of neural methods' advantages for prediction problems---that they are highly adaptable to many different problem spaces, varying both in more traditional hyper-parameters such as learning-rate and neural network width, but also in much-more nuanced choices such as optimization method or input structure. ``Neural Networks,'' despite the term's usage in many settings, is actually less of a single model and more a label placed on an entire class of iteratively-optimized models. The work's aim has been to use ``off-the-shelf'' machine learning methods to understand what empirical concerns exist when placing these models naively in an otherwise-recognizable econometric instrumental variables setting, but for a neural network, the off-the-shelf model is highly dependent on the problem at hand. Unfortunately, this advantage of MLPs and other Neural Networks makes a full grid-search of the hyper-parameter space intractable. This requires us to restrict the grid-space somewhat to create a tractable solution, while allowing the model a good shot at choosing the ``correct'' specification. This restriction potentially handicaps the neural model's flexibility, and may produce higher average out-of-sample loss than the full set of model specifications could potentially produce.

Second, neural networks are computationally expensive to train---each combination of hyperparameters must be trained separately over many iterations, and for most optimization procedures it is useful to utilize different re-orderings of the data-set to reach a satisfactory loss-minimizing point. Even for less-complex data such as ours with relatively few observations, cross-validating even 100 hyperparameter combinations over 1000 synthetic datasets leads to prohibitively long training periods given our computational resources.

Last, neural networks are sensitive to their initialization point, which is not the case for any of our other methods included in this analysis. Unlike most other methods, neural-class models can find values for one of a number of potential loss-minimizing local optima for its parameters, and while those local optima can perform similarly well, they do not necessarily produce identical predictions and can fail on different subsets of data in different ways.

These differences make analyzing how cross-validating a neural network dictates hyper-parameters given a reasonable loss function more interesting because the choices a five-fold cross-validation approach might make are indicative of how the most flexible model chooses a specification given different search spaces. For our cross-validation, we use a five-fold cross-validation procedure to match our other methods, and use mean-squared error as our loss function. We fix a few hyperparameters in place---we use no regularization on the weights, and use an ``Adam'' optimizer with it's out-of-box/off-the-shelf learning rate of .001. We trained all models over 40 epochs, and used a batch-size of 10 observations. One unusual step we take is to introduce a leaky rectified linear activation function (ReLU) activation function to connect hidden layers. This was done to prevent the model from suffering from ``dying weights'' which is when parameters accidentally force a large number of activations to inappropriately ``ignore'' activations due to $a(x) = 0$ for all or many values of $x$.\footnote{ReLU was tried initially, however using ReLU on our data seemingly led to a complete shutdown of the predictive power when we attempted it on our weaker-instrument setting. Even using dropout, the MLP's performance was poor in predicting out of sample. We cannot be certain that the dying weights problem was the cause of this issue, but adding a slight negative slope for activations less than 0 seems to have mitigated the behavior.}

We began by creating three separate hyper-parameter search spaces distinguished by maximum-allowable width and depth. The 'shallow/wide' neural networks are allowed to choose from hidden-layer representations that are 16, 32, 64, 256, and 512 nodes in hidden layer width respectively. This model is then restricted to contain at most a single hidden layer, but is also allowed to choose from a model that maps inputs to outputs directly, using a linear activation function. This functional form is sensible given x is linear in z for our DGPs. Excluding the 0-hidden layer case would prevent the cross-validation procedure from finding the easiest approximation for a linear functional form. The cross-validation procedure allows differences in regulation by choosing between a dropout rate of .1 or .2.

The ``narrow/deep" neural network is instead allowed to choose from a representation with two, three, four, or five hidden layers each with a number of nodes (width) equal to 16, 32, or 64. This model, too, is allowed to choose from the simple linear mapping of inputs to output, for the same reasons as above.

The last search space, referred to as the ``unrestricted'' neural network is allowed to choose from any combination of the hyperparameters offered to the narrow or shallow networks---from zero through five hidden layers and using the full complement of widths. 

To get a sense of how these search-spaces choose models on average, these three search spaces were used to cross-validate over our first 25 datasets, from which we generated a full list of 125 folds. These cross-dataset folds were then used to find, for each search space, the average out of sample MSE across all 125 folds.

From the set of available models available to every search space and for each iteration, we chose two models at random weighted by their average out of sample MSE. These probabilities were chosen using the weighted upper-tail normal CDF, normalized such that all weights for a given search space sum to one. Formally, where $i$ is a given set of hyperparameters, $j$ is a search-space and $\mu_j$ is the mean out of sample MSE for a given search space:

\begin{align}
p^{chosen}_{i,j} = \int_{z^{mse}_{i,j}}^{\infty} \frac{1}{\sqrt{2 \pi}}e^{- \frac{z}{2}} dz~,
\end{align}
\begin{align}
z^{mse}_{i,j} := \frac{mse_{i,j} - \mu_{j}}{se(mse_j)}\label{eq:a3}~.
\end{align}

For each iteration and using the probabilities above, two models are chosen at random for each search space, and then cross-validated again using a 5-fold procedure. The ``winning'' model is chosen by lowest out-of-sample MSE, and is used to predict $\hat{x}$ for the first stage.

For a visual explanation and overview of the results from the selection method, see Figure \ref{fig:nn-depth}.

\subsubsection{Low-bias methods}

Because machine learning algorithms are designed to minimize loss (maximizing fit), the fact that $\text{Cov}(\hat{x},e)\neq 0$ is partially by design. To see this fact, consider any prediction method that minimizes mean-squared error (MSE)---conditional on the training data $\{x,\mathbf{z}\}$:

\begin{align}
    \text{MSE}(\hat{x},x | x,\mathbf{z})
    &= \underbrace{\bigg( x - \expect*{\hat{x} | x,\mathbf{z}} \bigg)^2}_{\left(\text{Bias of } x \text{ for } \hat{x}\right)^2} + \underbrace{\expect*{~\left(\hat{x} - \expect*{\hat{x} | x,\mathbf{z}}\right)^2 | x,\mathbf{z}~}}_{\text{Cond. } \text{Var}(\hat{x})~} + \underbrace{\expect*{~\varepsilon^2 | x,\mathbf{z~}}~}_{\text{Cond. }\text{Var}(\varepsilon)}
    \label{eq:mse-decomp}
\end{align}
where $\varepsilon$ is the irreducible error---the unknowable disturbance from the DGP of $x$, \ie $x=f(\mathbf{z}) + \varepsilon$.\footnote{These expectations are conditional on the given dataset; $\varepsilon$ and $\hat{x}$ are conditionally independent by definition. The expectation term is conditional on data observed, so for simplicity, the term $E_D(\hat{x}(z;D))$ will simply be referred to as $\hat{x}$.}

Equation (\ref{eq:mse-decomp}) highlights that in an MSE-minimization problem, $\hat{x}$ is the only component of MSE that a learning algorithm can change ($x$, $\mathbf{z}$, and $\varepsilon$ are all data dependent). This fact leads to the widely discussed variance-bias tradeoff---an arbitrary estimator will generally face a negotiate between low-variance predictions and low-bias predictions. Many traditional econometric estimators result from prioritizing zero bias and then selecting the minimum-variance estimator from this class of unbiased estimators. As (\ref{eq:mse-decomp}) points out, these estimators could reduce their out-of-sample MSE by \textit{accepting} some bias and reducing variance. This tradeoff is at the heart of the prediction improvement many out-of-the-box algorithms offer relative to plain OLS.

\subsection{Appendix: Neural approaches to measuring causal response}


Neural networks and their offspring offer just such a route to explore data that falls outside of the traditional bounds, whether that is to use transformers for text data, or convolutional neural networks (CNNs) for imaging. However, many of the same problems apply to these tools as were outlined in the paper. In order to make full use of them in a two stage approach, the same stringent restrictions are required to generate meaningful and unbiased coefficients in a two-stage framework. Indeed, when running a cross-validated feed-forward network to produce a meaningful first stage estimation with our high-complexity data most simulations with any hidden layers simply reproduced an approximation of $\beta$ close to that of naive OLS.\footnote{See Appendix Section \ref{sec:MLP} for full details of the MLP methods.} There is a burgeoning field of research in machine learning that strives to understand IV problems under less-parametric (though generally still somewhat parametric) causal structures and these methods are seeing success in both simulated and real-world data. The downside to using these powerful methods is that they require a new framework in which to understand them, and make interpretation of treatment effects more challenging. 

The first of the recent batch of machine learning instrumental variables papers is referred to as ``Deep IV'' \cite{Hartford2017}. The authors throw away the linear functional form for $x = f(z,u)$ in the ``first stage,'' but assume linearly additive confounding variables and learn the causal structure with a two-part one-pass neural network model. The authors do this by recasting the econometric approach to instrumental variables into two interlinked problem spaces - estimating the conditional distribution $g(x|z)$ and then using the approximation of $x$ given $z$ to predict $y$. This creates difficulties because such methods produce good counterfactual predictions, but have a harder time matching the clean interpretable causal effect of X on y when compared to traditional econometric approaches. Further, because of the flexibility in functional form, models of this category tend to have more trouble outside of $supp(z_{train})$ or $supp(g(x|z_{train}))$ that are observed in a training sample - and it's difficult to apply a post-analysis structure to such a model to gather understanding on counterfactuals where $z_{test}$ is considerably different than $z_{train}$. Further, many other methods have been created and can be used to estimate instrument-identified causal effects using a similar semi-parametric two-stage function that can identify complex functional forms in either first or second stages \cite{Bennett2020} and \cite{xu2020} and improve on edge-of-support marginal effects.\footnote{The methodology contained in \cite{xu2020} is particularly useful, because it does not predict $X$ directly, but rather applies Neural Networks to the task of learning polynomial forms to pass through first and second stages in a 2SLS (with an L2 penalty) and may mostly avoid components a and b as described earlier.} Both of these papers and Deep IV are able to produce causal inferences using images as instruments---something that a regression would not be able to meaningfully do without some form of pre-model dimensionality reduction.

Neural approaches to causal inference are also not limited to use semi-parametric structural forms for heterogenous treatment effects. \cite{kilbertus2020class} created a neural network to identify the total set of conditional causal effects given a fully non-parametric instrumental variable analysis. (With the very reasonable assumption placed on the function of unobserved noise that it does not feature infinite discontinuities, for example.) 

In spite of the massive technical improvements these models have made, trying to extract a beta-equivalent from the existing models is difficult, though interpretable machine learning methods do exist. Unfortunately, extracting meaningful information using prediction-explanation methods such as \cite{Ribeiro2016} about how a result is produced, or to infer what kind of economic information can be gathered from the weights within a model, neural networks are hard to interpret \cite{Wang2019}. This makes comparing such models to traditional 2SLS or econometric approaches for ATE approximation difficult---and produces complications in choosing benchmarks as to how to evaluate them.

\subsection{Appendix: Complications of ML and monotonicity}
\label{sec:Monotonicity}

One often overlooked assumption in instrumental variables irrelevant under the assumption of constant treatment effects, but, without guaranteed constant treatment effects, is referred to as 'monotonicity' of heterogenous treatment effects. If a researcher is willing to simply find the 'Local Average Treatment Effects' or LATE \cite{Angrist2009}, 2SLS can under certain circumstances recover that estimate. Simply put, this means that while treatment effects can vary across our population, response of endogenous variable $x_i$ to instrument $z_i$ must move in the same direction for all individuals $i$. For canonical cases of binary instruments and treatments, this boils down to an assumption of 'no defiers'.

Our results as written do not conclude one way or another about the implications of the monotonicity assumption with regards to the machine learning in the first stage, as our primary datasets feature homogenous treatment effects are constant and equal to $\beta$. However, relaxing from constant treatment effects to continuous, monotonicity-preserving and heterogenous treatment effects, under some circumstances using nonlinear methods can lead to inefficiency in estimates of the LATE.

To illustrate a very simple example, imagine a case estimating the coefficient $\beta_1$ where $x_i = 1 + \sum_{v=1}^7z_i*\gamma_i + \varepsilon_i + u_i$, where $\gamma_i \sim \gamma(.5,4)$, thus $E(\gamma_i) = 2$, $min(\gamma_i) = 0$, $Y = 1 + X_i\beta_1 + u_i + \epsilon_i$ and all error terms $\epsilon_i, \varepsilon_i, u_i ~ N(0,1)$. This extends the strengthened exclusion restriction from $E(z|\varepsilon) = 0$ to $E(z|\varepsilon) = E(z|\eta X) = 0$ where $\eta = \gamma_i - E(\gamma_i)$.\footnote{See \cite{Heckman2006} for the original treatment of $\eta$ in the IV case.} In this case, the instruments are not themselves weak, but have weak effects on varying members of the sample. If a machine learning algorithm conditions its predictions on the instruments directly, and the coefficients $\gamma_i$ are sufficiently varied, $\hat{x_{ssml}}$ will only be a better estimate in expectation.

In this case; under the stronger exclusion restriction described above and using a MLSLS strategy as described in \cite{Chen2020-IV} with a non-linear meta-model, will produce an unbiased result, but also may result in inflated standard errors relative to a 2SLS or SSIV approach. This is because, under this kind of monotonicity, linear IV becomes a weighted average of marginal treatment effects.\footnote{which the reader should turn to \cite{Heckman2005} and \cite{Heckman2006} to see a formal treatment of.} If the econometrician believes this type of variation exists, the assumptions required for 2SLS' validity are strong, and MLSLS' are stronger. ML makes no guarantees on recovering this weighted sum in the same manner, so there exists important future work to examine how exactly this may impact structural estimates of $\beta$.

\end{appendices}


\end{document}